\documentclass[5p]{elsarticle}

\usepackage{amssymb}
\usepackage{amsmath}
\usepackage[utf8]{inputenc}

\usepackage{algorithmic}
\usepackage[ruled,vlined]{algorithm2e}

\usepackage{caption}
\usepackage{subcaption}

\usepackage{hyperref}
\hypersetup{
	colorlinks=true,
	linkcolor=blue,
	allcolors=true
}

\usepackage{nomencl} 

\makenomenclature

\usepackage{libertine}
\usepackage{libertinust1math}

\journal{International Journal of Electrical Power \& Energy Systems}

\begin{document}

\makeatletter
\def\ps@pprintTitle{%
 \let\@oddhead\@empty
 \let\@evenhead\@empty
 \def\@oddfoot{\it Postprint}%
 \let\@evenfoot\@oddfoot}
\makeatother

\begin{frontmatter}

	\title{Topology-based Approximations for $\mathcal{N}-1$ Contingency
	Constraints in Power Transmission Networks}

	\author[1]{Amin Shokri Gazafroudi\corref{cor1}}
	\ead{shokri@kit.edu}
	\cortext[cor1]{Corresponding author}

	\author[1,2]{Fabian Neumann}
	\ead{f.neumann@tu-berlin.de}

	\author[2]{Tom Brown}
	\ead{t.brown@tu-berlin.de}

	\address[1]{Institute for Automation and Applied Informatics, Karlsruhe Institute of Technology (KIT), 76131 Karlsruhe, Germany}

	\address[2]{Institute of Energy Technology, Technical University of Berlin, Einsteinufer 25, 10587 Berlin, Germany}

	\begin{abstract}

		It is crucial for maintaining the security of supply that transmission
		networks continue to operate even if a single line fails. Modeling
		$\mathcal{N} - 1$ security in power system capacity expansion problems
		introduces many extra constraints if all possible outages are accounted
		for, which leads to a high computational burden. Typical approaches to
		avoid this burden consider only a subset of possible outages relevant to
		a given dispatch situation. However, this relies on knowing the dispatch
		situation beforehand, and it is not suitable for investment optimization
		problems where the generation fleet is not known in advance. In this
		paper, we introduce a heuristic approach to model the fully secured
		$\mathcal{N}-1$ feasible space using a smaller number of constraints in
		a way that only depends on the topology of transmission networks. In our
		proposed approach, the network's security is modelled by comparing the
		polytope of the feasible space of nodal net power obtained from the
		security-constrained linearized AC optimal power flow problem. To
		approximate this polytope, a buffer capacity factor is defined for
		transmission lines in the $\mathcal{N}-0$ secure case, thereby avoiding
		the introduction of many additional constraints. In this way, three
		approaches are introduced for obtaining a buffer capacity factor
		consisting of approximate, robust and line-specific approaches. Finally,
		the performance of our proposed approaches is assessed in different
		scales of transmission networks for determining the proposed buffer
		capacity factors, contingency analysis and economic evaluation.
		Moreover, we find that our proposed heuristics provide excellent
		approximations of the fully secured $\mathcal{N}-1$ solutions with a
		much lower computational burden.

	\end{abstract}


\end{frontmatter}

\begin{footnotesize}
	\printnomenclature[1.5cm]
\end{footnotesize}

\setlength{\nomlabelwidth}{1.5cm}

\section{Introduction}

\subsection{Motivation}

Security and reliability are defined as two of the most important goals of power
systems to supply power demand continuously \citep{schnyder1988integrated}.
Thus, $\mathcal{N}-1$ criteria have been introduced to retain the power system
in a secure state when a single outage of power system components occurs
\citep{glanzmann2006incorporation}. Conventionally, $\mathcal{N}-1$ security is
modelled by many extra constraints to account for all outages, which leads to a
high computational burden, even when the flow physics is linearized (a common
approach for operational and capacity expansion problems). Typical approaches to
avoid this burden are to consider only a subset of possible outages which are
relevant for a given dispatch situation, iteratively adding the outage
constraints until the network can be demonstrated by experimentation to be
$\mathcal{N}-1$ secure \citep{tejada2017security}. This approach depends on the
dispatch situation and is unsuitable for investment optimization, where many
dispatch situations are considered simultaneously and the generation is not
known in advance. In the following section we summarise different approaches in
the literature for modelling security constraints, so-called
security-constrained linear optimal power flow (SCLOPF), to guarantee the power
grid's reliability and prevent line overloading and cascading failures in the
system.

\subsection{Related Works}

Authors in \citep{glanzmann2006incorporation,hug2009n} presented how the current
injection method retains the power system's security in the optimal power flow
problem by compensating power flow of the failed line based on injecting the
virtual currents to prevent  overloading in rest of the lines in the
transmission network. Authors in \citep{hedman2009optimal} presented a
mixed-integer program for the linearized AC optimal dispatch problem with
transmission switching considering $\mathcal{N}-1$ constraints. In
\citep{hedman2009optimal}, it has been evaluated how transmission switching can
keep the security of the power system based on $\mathcal{N}-1$ criteria. The
transmission switching problem was also considered by authors in
\citep{hedman2010co,zhang2018stochastic}. A joint unit commitment and
transmission switching problem considering $\mathcal{N}-1$ constraints has been
presented in \citep{hedman2010co}. A bi-level stochastic optimal switching model
has been introduced for determining line switching strategy considering
$\mathcal{N}-1$ limitations in \citep{zhang2018stochastic}. A preventive SCOPF
problem has been proposed in \citep{dong2012practical} where authors assessed
the performance of their proposed model in planning and operation problems of
the power system. In the preventative SCOPF problems, the proposed problem
consists of decision-making variables which are feasible for both
pre-contingency (before single line outages) and post-contingency (after single
line outages), simultaneously \citep{hinojosa2018preventive}. Ref.
\citep{karoui2008large} proposed a risk-based decomposed model for contingencies
in the power system. In \citep{capitanescu2019approaches}, contingencies
relaxation has been proposed. Authors in \citep{li2019security} presented a
security-constrained multi-objective optimal power flow method considering
$\mathcal{N}-1$ security constraints. Ref. \citep{fan2012n} proposed a
mixed-integer programming model for linear optimal power flow considering
$\mathcal{N}-1$ conditions for preventing cascading failures in the power
network. In \citep{weinhold2020fast}, an algorithm has been proposed for
reducing the constraints introduced by contingency scenarios of the SCOPF
problem and finding the feasible space for nodal power injected based on a
geometric algorithm. In \citep{poyrazoglu2015optimal}, authors introduced an
optimal topology control model considering $\mathcal{N}-1$ constraints in the
power systems. In \citep{ardakani2013identification}, an umbrella constraint
discovery problem has been introduced where the SCOPF problem is approximated by
the linearised AC power flow. Authors in \citep{aravena2020transmission}
proposed a framework for modeling zonal electricity markets and developed
cutting-plane algorithms for considering $\mathcal{N}-1$ security conditions.

\subsection{Contributions}

Although various models have been introduced in the literature to model
$\mathcal{N}-1$ security conditions in different applications of power systems,
most of them are dependent on properties of components of the system, e.g.
generators and loads, which increases the computational burden beyond what is
solvable in reasonable time in the context of multi-period capacity expansion
models. Moreover, there exist numerous alternative approaches with regards to
contingency conditions in light of new technologies such as large-scale energy
storage systems, e.g. batteries, that can be placed at the connecting nodes of
transmission lines outage which can compensate as \textit{virtual lines} for
some time. In this paper, we present a topology-based $\mathcal{N}-1$ security
method for transmission line outages in a preventive SCOPF problem that can be
applied in power system expansion problems. In other words, the power system's
security is solely studied based on the topology of the transmission network
without relying on information about generators, loads and energy storage
systems. Thus, this paper proposes a heuristic model for $\mathcal{N}-1$
security constraints of the transmission grid by adding buffer capacity factor
for transmission lines. In this way, we propose three heuristic approaches,
consisting of robust, approximate and line-specific, without adding constraints
for every single line outage. Our proposed models for $\mathcal{N}-1$ security
constraints are based on a $\mathcal{N}-0$ secured linear optimal power flow
formulation by adding a buffer capacity factor for each transmission line. For
instance, the loading of a line may be limited to 70\% of its real thermal
capacity as an industry-standard
\citep{brown2016optimising,schlachtberger2017benefits}. This way, the remaining
30\% of the thermal capacity is kept as a reserve to guarantee security and
reliability if another line fails \citep{siemens}. However, the approximation
has never been justified or shown to be robust, nor have line-specific factors
been developed. Using a buffer capacity factor on the line capacities instead of
the full set of $\mathcal{N}-1$ constraints saves many heavy constraints and
non-zero values in the constraint matrix. However, a price is paid for the lower
computational burden as the heuristic approaches presented herein may be too
optimistic or too conservative. The main contributions of this paper can be
summarized as follows:

\begin{itemize}
	\item We propose a heuristic approach for $\mathcal{N}-1$ security
	constraints of the transmission power grid based on adding a buffer capacity
	factor for transmission lines that dramatically reduces number of necessary
	constraints.
	\item We investigate three approaches to find the buffer capacity factor for
	all or individual transmission lines.
	\item The performance of the fully secured $\mathcal{N}-1$ network is
	compared with our proposed heuristic approaches.
	\item A sensitivity analysis of the proposed approaches is performed in
	different network topologies and German transmission networks.
\end{itemize}

The rest of this paper is organized as follows. Section \ref{sec:formul}
presents the problem formulation for the linearized AC optimal power flow
problem under $\mathcal{N}-1$ security constraints. In Section \ref{sec:method},
our heuristic approaches to model $\mathcal{N}-1$ secure network are introduced.
Then, the simulation results are discussed in Section \ref{sec:simulation}.
Finally, our findings are summarised in Section \ref{sec:conclusion}.

\section{Problem Formulation}
\label{sec:formul}

\subsection{Linearized AC Optimal Power Flow Problem}

We consider the linearized AC optimal power flow (LOPF) problem that seeks to
minimize the operational cost of generator dispatch under network constraints
with fixed capacities. Nonlinear and non-convex formulations of power flow have
not been considered in this paper because they cannot be solved within
reasonable time in the context of large capacity planning problems. The problem
is classified as a linear problem (LP) and no capacity expansion of generation,
storage or transmission infrastructure is assumed. The objective function for
the LOPF problem is given by (\ref{eq:lopf_obj}).
\begin{equation}\label{eq:lopf_obj}
	\min_{g_{s,t},f_{\ell,t}} \sum_{s,t} o_s g_{s,t},
\end{equation}
where $o_s g_{s,t}$ represents operation cost for generator $s$ at time $t$.
This way, $g_{s,t}$ represents the dispatched power of generator $s$ at time
$t$. The LOPF problem is constrained by physical equations and constraints which
are presented in the following. Eq. \eqref{eq:lopf_bal} presents the balancing
constraint based on Kirchhoff's Current Law (KCL)  requiring matching generation
and demand at bus $i$ and time $t$
\begin{equation}\label{eq:lopf_bal}
	\sum_s M_{i,s}g_{s,t} - d_{i,t} = \sum_\ell K_{i,\ell}f_{\ell,t}, \quad \forall i, \forall t
\end{equation}
where $M_{i,s}$ maps generator $s$ at bus $i$. In other words, $M_{i,s}=1$ if
generator $s$ is located at bus $i$, otherwise $M_{i,s}=0$. Besides, $d_{i,t}$
represents electricity demand at bus $i$ and time $t$, and $f_{\ell,t}$ states
power flow in line $\ell$ and time $t$. Moreover, $K_{i,\ell}$ represents the
incidence matrix between buses and lines as represented by \eqref{eq:incidenc K}
\begin{equation}\label{eq:incidenc K}
	K_{i,\ell} =
	\begin{cases}
		1, \quad \hspace{0.15 cm}  \text{if bus $i$ is started bus of line $\ell$,} \\
		-1 \quad \text{if bus $i$ is ended bus of line $\ell$,}                     \\
		0 \quad \hspace{0.25 cm} \text{if bus $i$ is not connected to line $\ell$.}
	\end{cases}
\end{equation}
Moreover, maximum and minimum limitations of dispatched power of generators are
expressed by \eqref{eq:lopf_gen2}
\begin{equation}\label{eq:lopf_gen2}
	0 \leq g_{s,t} \leq \overline{G}_{s,t}, \quad \forall s, \forall t
\end{equation}
where $\overline{G}_{s,t}$ represents the time-dependent potential of power
generation for generator $s$ at time $t$. Also, Eq. \eqref{eq:lopf_co2} states
the $CO_2$ constraint for avoiding  $CO_2$ emission exceeds a desired target
level.
\begin{equation}\label{eq:lopf_co2}
	\sum_{s,t} \frac{e_s}{\eta_s} g_{s,t} \leq \mathcal{E}_{CO_2},
\end{equation}
where $e_s$ is the specific emission of the fuel, $\eta_s$ is the generator
efficiency, and $\mathcal{E}_{CO_2}$ is a fraction of the $1990$ emissions in
electricity \citep{PyPSA}.

\subsection{Power Flow Constraints}

In this section, nodal and line-specific power flow equations and constraints of
the LOPF problem are represented. Power nodal balancing is one of the
fundamental equations in the power system. In this paper, power loss is ignored
for simplicity. Then, we obtain
\begin{equation}\label{eq:line_balance}
	\sum_{i=1}^{\mathcal{B}} p_{i,t} = 0,
\end{equation}
where $p_{i,t}= \sum_s M_{i,s}g_{s,t} - d_{i,t}$, and it represents net power
injection at bus $i$ and time $t$. Thus, positive (negative)  value of $p_{i,t}$
presents network generation (load). Here, bus $i=1$ is chosen as a slack bus in
the grid. In other words, $p_{2,t}, \dots, p_{\mathcal{B},t}$ are independent
values and $p_{1,t}$ is the dependent value at each time snapshot. Authors in
\citep{wood2013power,horsch2018linear,hinojosa2016improving,hinojosa2017stochastic}
present the relation between power flow and nodal net power is given by
\eqref{eq:ptdf}
\begin{eqnarray}\label{eq:ptdf} f_{\ell,t}=\sum_{i=2}^{\mathcal{B}}
	PTDF_{\ell,i}p_{i,t},\quad\forall \ell,\forall t.
\end{eqnarray}
where $PTDF_{\ell,i}$ represents power transfer distribution factor (PTDF)
between line $\ell$ and bus $i$ which is given by $PTDF=BK^{T}\Lambda^{*}$.
Moreover, $B$ is a diagonal matrix of branch susceptance as it arrays are given
by \eqref{eq:B_array}
\begin{equation}\label{eq:B_array}
	B_{\ell,k} =
	\begin{cases}
		\frac{1}{x_{\ell}}, \quad \text{if } \ell = k \\
		0 \quad \quad \text{otherwise}
	\end{cases}
\end{equation}
where $x_{\ell}$ is a reactance for line $\ell$. Moreover, $\Lambda^*$ is a
pseudo-inverse matrix of $\Lambda$, as nodal susceptance matrix, which is
obtained by $\Lambda = KBK^T$. Thus, as $i = 1$ is assumed as a slack bus, then
$PTDF_{\ell,1}=0$. Maximum and minimum constraints of power flow are represented
in (\ref{eq:line_max})
\begin{eqnarray}\label{eq:line_max} - F_{\ell} \leq f_{\ell,t}\leq F_{\ell},
	\quad\forall \ell,\forall t
\end{eqnarray}
where $F_{\ell}$ is the capacity of line $\ell$. To model the $\mathcal{N}-1$
secure network, it is necessary to define the $\mathcal{N}-0$ secure network.
The $\mathcal{N}-0$ secure network is only secure as long as there is no line
outage. According to $\mathcal{N}-0$ security conditions, Eq.
(\ref{eq:ptdf_max}) is obtained based on (\ref{eq:ptdf}) and (\ref{eq:line_max})
\begin{eqnarray}\label{eq:ptdf_max} - F_{\ell} \leq \sum_{i=2}^{\mathcal{B}}
	PTDF_{\ell,i}p_{i,t}\leq F_{\ell}, \quad\forall \ell,\forall t.
\end{eqnarray}
In this way, $\mathcal{B}-1$ independent nodal net powers ($p_{2,t}, \dots,
p_{\mathcal{B},t}$) make the $\mathcal{B}-1$ dimensional space at each time
snapshot based on (\ref{eq:ptdf_max}). In other words, at time $t$, Eq.
(\ref{eq:ptdf_max}) constrains the $(\mathcal{B}-1)$--dimensional polytope with
at most $2\mathcal{N}$ faces where $\mathcal{N}$ is number of transmission lines
in the network. We define this as the $\mathcal{N}-0$ secure polytope
$Q^{\mathcal{N}-0}$.

\subsection{Security Constraints}

In this section, security constraints are presented that guarantee no
overloading occurs in transmission lines if there is a single line outage
($\mathcal{N}-1$ criteria). In this way, if line $k$ fails, the flow in line
$\ell$ at time $t$ after the outage of line $k$ ($f_{\ell,t}^{k}$) is related to
the flows before the line outage in lines $\ell$ ($f_{\ell,t}$) and $k$
($f_{k,t}$) based on the line outage distribution factor (LODF) represented in
Eq. (\ref{eq:line_outage}) \citep{ronellenfitsch2017dual}. Based on Eqs.
(\ref{eq:ptdf}) and (\ref{eq:line_outage}), Eq. (\ref{eq:lodf_1}) is obtained
\citep{ronellenfitsch2017dual}
\begin{eqnarray}\label{eq:line_outage}
	f_{\ell,t}^{k}=f_{\ell,t}+LODF_{\ell,k}f_{k,t}, \quad\forall \ell, \forall
	t\\
	\label{eq:lodf_1}
	f_{\ell,t}^{k}=\sum_{i=2}^{\mathcal{B}}[PTDF_{\ell,i}+LODF_{\ell,k}PTDF_{k,i}]p_{i,t},
	\forall \ell, \forall t
\end{eqnarray}
where the $PTDF$ and $LODF$ matrices are model coefficients and calculated
before outage of line $k$. Moreover $LODF$ is calculated according to Refs.
\citep{ronellenfitsch2017dual, guler2007generalized, guo2009direct} by Eq.
\eqref{eq:lodf}
\begin{eqnarray}\label{eq:lodf} LODF_{\ell,k}= \frac{[PTDF \cdot K]_{\ell,k}}{1-
	[PTDF \cdot K]_{\ell,\ell}}, \quad\forall \ell,
\end{eqnarray}
where $LODF_{\ell,k}=-1$ in case of $\ell=k$. If there is an outage in one line
(e.g. line $k$ fails), the power flow in lines after a single line outage is
limited their corresponding thermal capacity which is given by
(\ref{eq:l_k_max})
\begin{eqnarray}\label{eq:l_k_max} - F_{\ell} \leq f_{\ell,t}^{k}\leq F_{\ell},
	\quad\forall \ell, \forall t.
\end{eqnarray}
Additionally, according to (\ref{eq:lodf}) and (\ref{eq:l_k_max}), Eq.
(\ref{eq:lodf_max}) is obtained:
\begin{eqnarray}\label{eq:lodf_max} - F_{\ell} \leq
	\sum_{i=2}^{\mathcal{B}}[PTDF_{\ell,i}+LODF_{\ell,k}PTDF_{k,i}]p_{i,t}\leq
	F_{\ell}, \nonumber\\
	\quad\forall \ell, \forall t.
\end{eqnarray}

Thus, for a fully secured $\mathcal{N}-1$ network, both Eqs. (\ref{eq:ptdf_max})
and (\ref{eq:lodf_max}) must hold. This adds $2\mathcal{N}(\mathcal{N}-1)$
constraints from Eq. (\ref{eq:lodf_max}) to the $2\mathcal{N}$ constraints from
Eq. (\ref{eq:ptdf_max}) ($\mathcal{N}-0$ secure network) at each time snapshot.
In this way, at time $t$, the fully secured $\mathcal{N}-1$ network can be
represented by an $(\mathcal{B}-1)$--dimensional polytope $Q^{\mathcal{N}-1}$
consisting of $2\mathcal{N}^2$ constraints which is often computationally
infeasible for large networks that consider a wide range of operating
conditions. Thus, an approximation approach is needed to model $\mathcal{N}-1$
security conditions for large networks. Our proposed heuristic methods are
described in the following section.

\section{Methodology}
\label{sec:method}

In our proposed methodology, we approximate the shape of the $\mathcal{N}-1$
secure polytope $Q^{\mathcal{N}-1}$ by taking the $\mathcal{N}-0$ secure
polytope $Q^{\mathcal{N}-0}$ and introducing buffer capacity factors for each
line to constrain it further. In this approach, we modify Eq.
\eqref{eq:line_max} to:
\begin{eqnarray}\label{eq:l_C_max} - c_\ell F_{\ell} \leq f_{\ell,t}\leq c_\ell
	F_{\ell}, \quad\forall \ell, \forall t
\end{eqnarray}
where $c_\ell$ represents buffer capacity factor for line $\ell$ which is
between $0$ and $1$, and it expresses the share of line $\ell$ nominal capacity
that can be used. Thus, we define a new $(\mathcal{B}-1)$--dimensional polytope
$Q^{\textrm{heuristic}}$ with at most $2\mathcal{N}$ faces in the space of
possible nodal net powers. By definition, it is a subset of the $\mathcal{N}-0$
secure polytope $Q^{\textrm{heuristic}} \subset Q^{\mathcal{N}-0}$ as long as
$c_\ell \leq 1$ for all lines.

\subsection{Proposed heuristic approaches}
\label{sec:heuap}

We present three different approaches -- consisting of robust, approximate, and
line-specific -- to choose the buffer capacity factors to get the best match
between the polytope $Q^{\textrm{heuristic}}$ and the $\mathcal{N}-1$ polytope
($Q^{\mathcal{N}-1}$). While the same buffer capacity factor is allocated for
all lines in robust and approximate approaches, the specific amount of buffer
capacity factors are determined for each line in the line-specific approach.
Additionally,  while $Q^{\textrm{heuristic}}$ is entirely inside the
$Q^{\mathcal{N}-1}$ in the robust and line-specific approaches,
$Q^{\textrm{heuristic}}$ is not entirely inside the $Q^{\mathcal{N}-1}$ in the
approximate approach and the volume of $Q^{\textrm{heuristic}}$ is equal to the
volume of the $Q^{\mathcal{N}-1}$ in the approximate one. For contracting
$Q^{\mathcal{N}-1}$, firstly, we must shape $Q^{\mathcal{N}-0}$ and polytope of
each single line failure scenario of the network. In this way,
$Q^{\mathcal{N}-\ell_{1}}$ and $Q^{\mathcal{N}-\ell_{n}}$ present the polytopes
if only lines $\ell_{1}$ and $\ell_{n}$ are failed, respectively. Hence,
$Q^{\mathcal{N}-1}$ represents the common space between $Q^{\mathcal{N}-0}$ and
polytope of all single line failure scenario
\begin{eqnarray}\label{eq:qn-1_shape} Q^{\mathcal{N}-1}= Q^{\mathcal{N}-0}
	\cap_{i}Q^{\mathcal{N}-\ell_i}
\end{eqnarray}

\subsubsection{Robust approach}

The robust approach is the most conservative one. In this approach, the buffer
capacity factor is the same for all lines $c_\ell = c^R$, and $c^R$ is chosen so
that the resulting polytope $Q^R$ is entirely inside the polytope of the fully
secured $\mathcal{N}-1$ network, $Q^R \subseteq Q^{\mathcal{N}-1}$, as shown in
Fig. \ref{fig:diagram}. This guarantees that any nodal imbalance inside $Q^R$
will always be $\mathcal{N}-1$ secure. However, it may be over-constrained such
that there are more cost-effective dispatch solutions outside $Q^R$ that are
still $\mathcal{N}-1$ secure, i.e. there may be better solutions in the set
$Q^{\mathcal{N}-1} - Q^R$.

\subsubsection{Approximate approach}
\label{sec:approximate_algorithm}

\begin{figure}[!t]
	\centering
	\includegraphics[width=0.85\columnwidth]{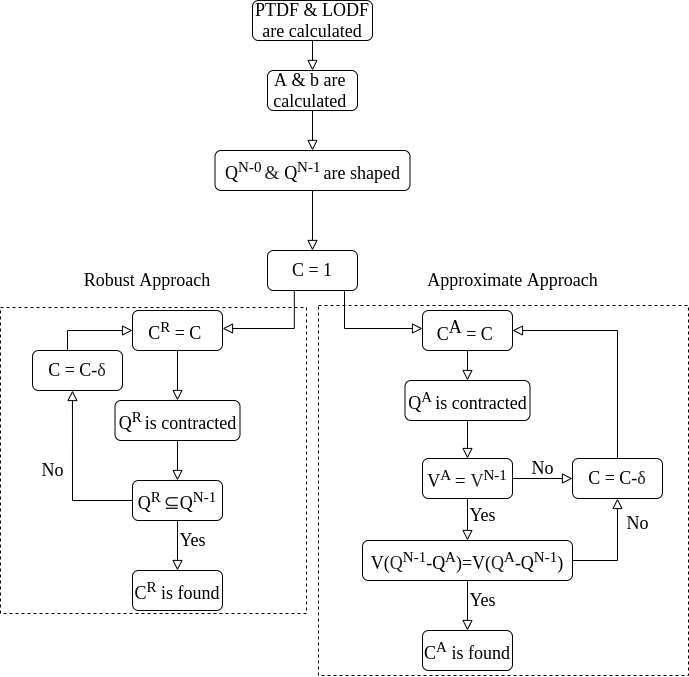}
	\caption{Step-by-step diagram of our proposed robust and approximate approaches to find buffer capacity factor for transmission lines.}
	\label{fig:diagram}
\end{figure}

This approach also has the same buffer capacity factor for every line, $c_\ell =
c^A$, but now $c^A$ is chosen such that the resulting polytope $Q^A$ has the
same volume $V^A$ as the fully secured $\mathcal{N}-1$ polytope
$Q^{\mathcal{N}-1}$, i.e. $V^A = V^{\mathcal{N}-1}$ which is less constraining
than the robust approach where $V^R \leq V^{\mathcal{N}-1}$. This means the
feasible space is the same size as the $\mathcal{N}-1$ polytope, in the hope
that the economic dispatch is better approximated by having a larger feasible
space than the conservative robust approach. Now, like the robust case, we may
still have $\mathcal{N}-1$ secure solutions that are outside $Q^A$, i.e. the set
$Q^{\mathcal{N}-1} - Q^A$ is non-trivial, but we may also have solutions in
$Q^A$ that are now no longer $\mathcal{N}-1$ secure, i.e. the set $Q^A -
Q^{\mathcal{N}-1}$ is non-trivial. Since $V^A = V^{\mathcal{N}-1}$, we have that
the volume of these two non-trivial spaces is equal as represented by Eq.
\ref{eq:vol_equal}. The step-by-step implementation diagram of our proposed
robust and approximate approaches are illustrated in Fig. \ref{fig:diagram}.
\begin{eqnarray}\label{eq:vol_equal} V(Q^{\mathcal{N}-1} - Q^A) = V(Q^A -
	Q^{\mathcal{N}-1})
\end{eqnarray}

\subsubsection{Line-specific approach}

This approach builds on the robust approach by using the freedom to set the
individual buffer capacity factor for each line, $c_{\ell}$, in a way that keeps
the overall polytope be inside $Q^{\mathcal{N}-1}$ while increasing the volume
of the polytope. Thus, the maximum power transfer capacity for each line is
determined based on the \textit{redundant capacity factor}, $t_{\ell}$,
according to the \textit{Edmond-Karp} algorithm
\citep{edmonds1972theoretical,witthaut2016critical}. This way, if $t_{\ell} >
1$, it represents that power flow above the nominal capacity of line $\ell$ does
not cause overloading in the rest of the lines in the network in the case of a
single line outage of line $\ell$. The redundant capacity factor $t_{\ell}$ for
line $\ell$ is calculated based on the LOPF problem in the network after (i)
removing line $\ell$ with start node $i$ and end node $j$, (ii) adding power
generation at node $i$ and load at node $j$ until one line becomes overloaded in
the transmission network, and vice versa. Based on $t_{\ell}$ and
$Q^{\mathcal{N}-1}$, $c_{\ell}$ is determined iteratively as presented in
Algorithm \ref{alg:c_line}. The algorithm starts with the robust polytope, then
increases $c_\ell$ for each line based on its corresponding $t_{\ell}$, then
incrementally reduces $c_\ell$ until the resulting polytope is robust, i.e.
within $Q^S \subseteq Q^{\mathcal{N}-1}$.

\begin{algorithm}[!t]
	\caption{Proposed algorithm to determine line-specific buffer capacity
	factor ($c_{\ell}$).}
	\begin{algorithmic}[1]
		\STATE Determine robust buffer capacity factor ($c^{R}$) \STATE
		Determine redundant capacity factor for each line ($t_{\ell}$) \STATE
		Update line capacity ($F_{\ell}^{'}\leftarrow t_{\ell}F_{\ell}$)
		\label{state:3} \STATE Determine $c^{R'}$ based on $F_{\ell}^{'}$ \STATE
		Set initial value for line-specific buffer capacity factor
		($c_\ell\leftarrow c^{R'}t_{\ell}$) \STATE Update value for
		line-specific buffer capacity factor\\ ($c_\ell\leftarrow \max (c^{R},
		\min(c_\ell,1) )$)
		\FOR{$l\leftarrow 1$ to $\mathcal{N}$}
		\WHILE{$Q^{\textrm{S}} \not\subset Q^{\mathcal{N}-1}$ and $c_{\ell}>
		c^{R}$}
		\STATE $c_{\ell}\leftarrow c_{\ell}-\delta ,\quad\forall \delta\in[0,1]$
		(e.g. $\delta=0.01$)
		\ENDWHILE
		\ENDFOR
	\end{algorithmic}
	\label{alg:c_line}
\end{algorithm}

\subsection{Finding the volume of the polytope}

As highlighted in Section \ref{sec:approximate_algorithm}, $c^A$ is found based
on comparison between $V^A $ and $V^{\mathcal{N}-1}$. To determine the volume of
the polytopes generated from our proposed approximate approach, a set $X$ of
random samples is used, where $X$ is a $\mathcal{B}$--dimensional vector and
represents $p_{i}$ for a single time. In this way, we use a quasi-random
low-discrepancy series, e.g. the \textit{Halton} sequence
\citep{kuipers2012uniform} to improve the convergence over pseudo-random
sampling. Thus, the space in the bounding box of Eqs. (\ref{eq:axb_max}) and
(\ref{eq:axb_min}) is covered more efficiently based on the $\mathcal{N}-1$
polytope defined by Eq. (\ref{eq:lodf_max}).
\begin{align}\label{eq:axb_max}
	\overline{X}  & = \max\{X\:|\:AX-b=0\}, \\
	\label{eq:axb_min}
	\underline{X} & = \min\{X\:|\:AX-b=0\},
\end{align}
where $A$ and $b$ are the left-hand side and the right-hand side of Eq.
(\ref{eq:lodf_max}), respectively. In this way,  $b$ is a $\mathcal{N}$--
dimensional vector and $A$ is a $\mathcal{N} \times \mathcal{B}$-- dimensional
matrix. Besides, $\underline{X}$ and $\overline{X}$ present lower and upper
bounds of $X$, respectively. Finally, samples constrained to (\ref{eq:axb}) are
selected for determining the volume of the polytope.
\begin{align}\label{eq:axb}
	AX \leq |b|,
\end{align}
where $x_{hit}$ is number of selected samples from $X$ and $x_{miss}$ is total
number of generated samples from $X$ in the bounding box and are not selected.
This way, the volume of the polytope is calculated based on a hit-miss
integration by Eq. \eqref{eq:vol}:
\begin{eqnarray}\label{eq:vol} V =
	\frac{(\overline{X}-\underline{X})x_{hit}}{x_{hit}+x_{miss}},
\end{eqnarray}
where $V$ represents volume of the polytope.

\subsection{Methodology for large numbers of buses}
\label{sec:large}

Our proposed heuristic approaches to find buffer capacity factors are based on
estimating volumes of polytopes. It is noteworthy that estimating volumes of
polytopes does not perform well on large-dimensional spaces because of the curse
of dimensionality. To circumvent this problem, we apply the algorithms proposed
in Section \ref{sec:heuap} on many smaller subsets of the buses separately.
Moreover, a large number of subsets of buses is necessary for accurate results,
which is a drawback for this method. For instance, for a network with 50 buses,
we can take a group of $\mathcal{B}_z$ buses, e.g. $5$, buses and explore the
$\mathcal{B}_z$--dimensional polytope defined by the space of possible
dispatches of these $\mathcal{B}_z$ buses, with all other buses having zero
power dispatch. Thus, all line constraints are still active, and this is
equivalent to only examining $\mathcal{B}_z$ columns of the matrix $A$ in
equation \eqref{eq:axb}, but keeping all the rows that represent all the line
constraints in the model as represented in (\ref{eq:A_cluster}).
\begin{equation}\label{eq:A_cluster}
	\begin{array}{r@{\,}l}
		    &
		\begin{matrix}
			\mspace{95mu}
			\overbrace{\rule{1cm}{0pt}}^{\mathcal{Z}}
			\mspace{35mu}
			\overbrace{\rule{1cm}{0pt}}^{\mathcal{Z}}
		\end{matrix}
		\\
		A = &
		\begin{pmatrix}
			a_{1,1}           & \cdots & a_{1,i}           & \cdots & a_{1,i'}           & \cdots & a_{1,\mathcal{B}}           \\
			\vdots            &        & \ddots            &        & \vdots             &        & \vdots                      \\
			a_{\ell,1}        & \cdots & a_{\ell,i}        & \cdots & a_{\ell,i'}        & \cdots & a_{\ell,\mathcal{B}}        \\
			\vdots            &        & \vdots            &        & \ddots             &        & \vdots                      \\
			a_{\mathcal{N},1} & \cdots & a_{\mathcal{N},i} & \cdots & a_{\mathcal{N},i'} & \cdots & a_{\mathcal{N},\mathcal{B}}
		\end{pmatrix}
	\end{array}
\end{equation}

In this way, the corresponding matrix $A$ for subset $\mathcal{Z}$ is given in
(\ref{eq:A_Z}).
\begin{equation}\label{eq:A_Z}
	A^{S}_{\mathcal{Z}} =
	\begin{pmatrix}
		a_{1,i}           & \cdots & a_{1,i'}           \\
		\vdots            & \ddots & \vdots             \\
		a_{\mathcal{N},i} & \cdots & a_{\mathcal{N},i'}
	\end{pmatrix}
\end{equation}

Moreover, matrix $A$ is represented as consisting of $A^{S}$ for all subsets.
\begin{equation}\label{eq:A_subset}
	A =
	\begin{pmatrix}
		A^{S}_{1} & A^{S}_{2} & \cdots & A^{S}_{\mathcal{Z}}
	\end{pmatrix},\quad\forall z.
\end{equation}

Geometrically this is equivalent to taking cross-sections of the polytope along
particular subsets of dimensions while setting the coordinates of other
dimensions to zero. By sampling many such subsets of nodes, corresponding to
different cross-sections of the polytope, we can build information on the whole
polytope. In this way, subsets can be selected by random selections or by
clustering buses that are geographically close to each other using k-means
algorithm \citep{kanungo2002efficient}. In this way, $c^R_z$ and $c^A_z$, which
are represented as robust and approximate buffer capacity factors for cluster
$z$, respectively, are determined according to robust and approximate approaches
described in Section \ref{sec:heuap}. According to the definition of $c^R$, we
have
\begin{eqnarray}\label{eq:crob} c^R = \min\{c_\ell \} , \quad\forall \ell.
\end{eqnarray}
As presented in Eq. (\ref{eq:crob}), the $c^R$ for the system must be the
minimum of the line-specific buffer capacity factors of all lines. In this way,
if we split the network to $\mathcal{Z}$ subsets, subset $z$ contains its
corresponding lines and robust buffer capacity factor, $c_{z}^{R}$, which is the
minimum of line-specific buffer capacity factor for lines belonged to subset
$z$. Thus, the $c^R$ for the network split to several clusters can be defined as
the minimum of robust buffer capacity of all subsets as represented in
(\ref{eq:crob_z}).
\begin{eqnarray}\label{eq:crob_z} c^R = \min \{c_{z}^{R} \} , \quad\forall z.
\end{eqnarray}

However, for determining $c^A$, we need to figure out the relation between the
buffer capacity factor and  the volume of $Q^A_z$ as compare to the volume of
$Q^{\mathcal{N}-1}_z$. In this paper, we define a criteria so-called volume
ratio ($\Upsilon_z$) for subset $z$, which is given in (\ref{eq:volumratio}).
Regarding to the definition of the approximate approach in Section
\ref{sec:heuap}, one of the conditions for finding $c^A$ for the whole polytope
is $\Upsilon = 1$. In this way, we propose a  criterion which is called the
common volume ratio, $\Psi$, as represented in (\ref{eq:cvc}). Thus, $c^A$ is
chosen to minimise the differences of the individual volume ratios from unity,
see (\ref{eq:C_A_de}).

\begin{eqnarray}\label{eq:volumratio}
	\Upsilon_z=\frac{V(Q^A_z)}{V(Q^{\mathcal{N}-1}_z)},\quad\forall z.\\
	\label{eq:cvc}
	\Psi(c) = \sum_z(\Upsilon_z(c)-1)^2,\quad\forall c.\\
	\label{eq:C_A_de}
	c^{A}=c,  \hspace{0.3 cm}\min_c \Psi(c).
\end{eqnarray}

An alternative method for finding line-specific capacity factors ($c_\ell$) is
proposed for large-scale networks based on the method for finding the robust
subset factor ($c^R_z$). In this way, the network is split to $\mathcal{N}$
subsets in which buses $i$ and $j$ belong in the same subset if they connect
directly with a transmission line. This way, matrix $A^{S}_{\ell}$ for the
subset of line $\ell$ where is ended to buses $i,j$ as represented in
(\ref{eq:A_line})
\begin{equation}\label{eq:A_line}
	A^{S}_{\ell} =
	\begin{pmatrix}
		a_{1,i}           & a_{1,j}           \\
		\vdots            & \vdots            \\
		a_{\mathcal{N},i} & a_{\mathcal{N},j}
	\end{pmatrix},\quad\forall  \hspace{0.1 cm} |K_{i,\ell}|\cdot|K_{j,\ell}|=1 .
\end{equation}

In other words, each subset consists of only two buses that are connected with a
transmission line and its line-specific buffer capacity factor, $c_\ell$, equals
the robust buffer capacity factor for the corresponding subset, $c^R_{z=l}$, as
seen in (\ref{eq:C_l_de})
\begin{eqnarray}\label{eq:C_l_de} c_\ell=c^R_z \hspace{0.3 cm},\quad\forall
	\hspace{0.1 cm} |K_{i,\ell}|\cdot|K_{j,\ell}|=1.
\end{eqnarray}

\begin{figure}[!t]
	\centering
	\includegraphics[width=0.7\columnwidth]{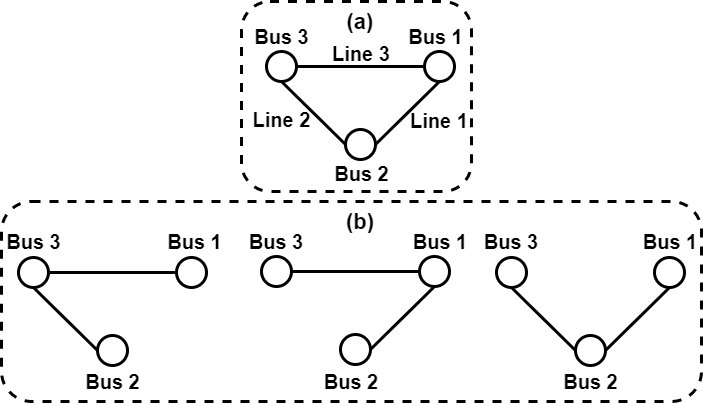}
	\caption{3-bus triangular network (a) $\mathcal{N}-0$ scenario (b)
	$\mathcal{N}-1$ scenarios.}
	\label{fig:node_n0_n1}
\end{figure}

\begin{figure}[!t]
	\begin{subfigure}{.4\textwidth}
		\centering
		\includegraphics[width=\linewidth]{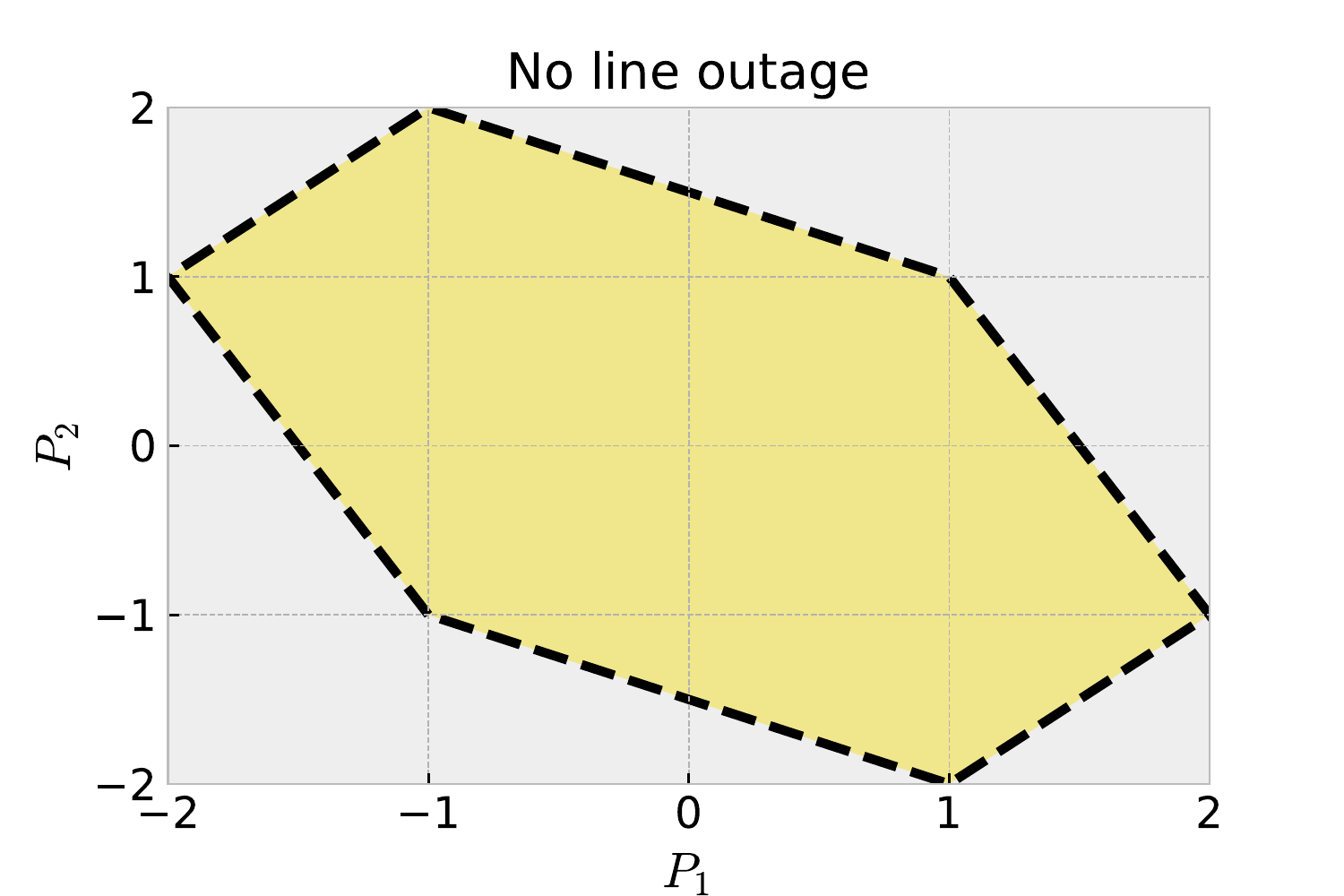}
		\caption{$\mathcal{N}-0$ secure network scenario}
		\label{fig:sub-first}
	\end{subfigure}
	\begin{subfigure}{.4\textwidth}
		\centering
		\includegraphics[width=\linewidth]{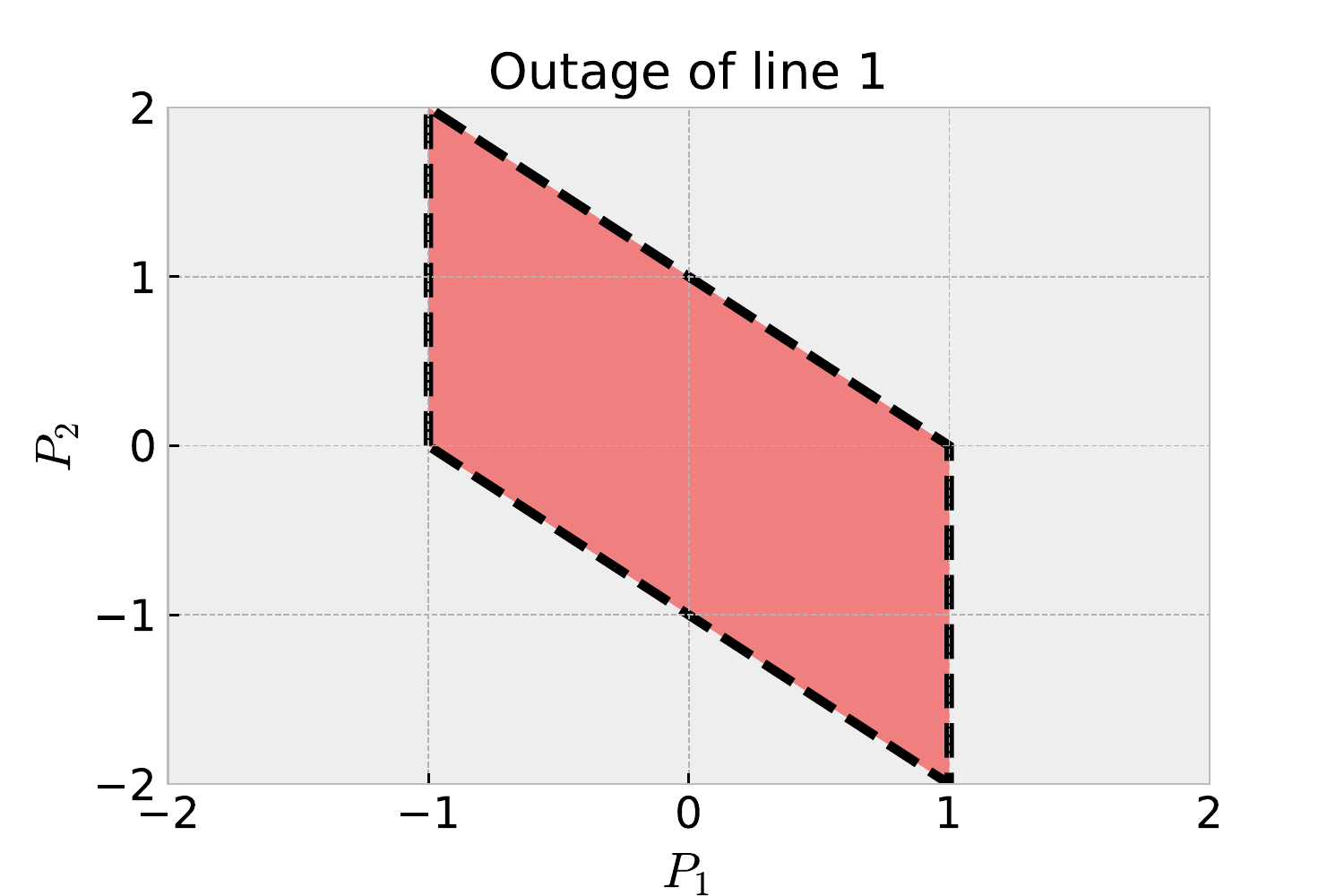}
		\caption{Line 1 failure scenario}
		\label{fig:sub-third}
	\end{subfigure}
	\vspace{2mm}
	\newline
	\begin{subfigure}{.4\textwidth}
		\centering
		\includegraphics[width=\linewidth]{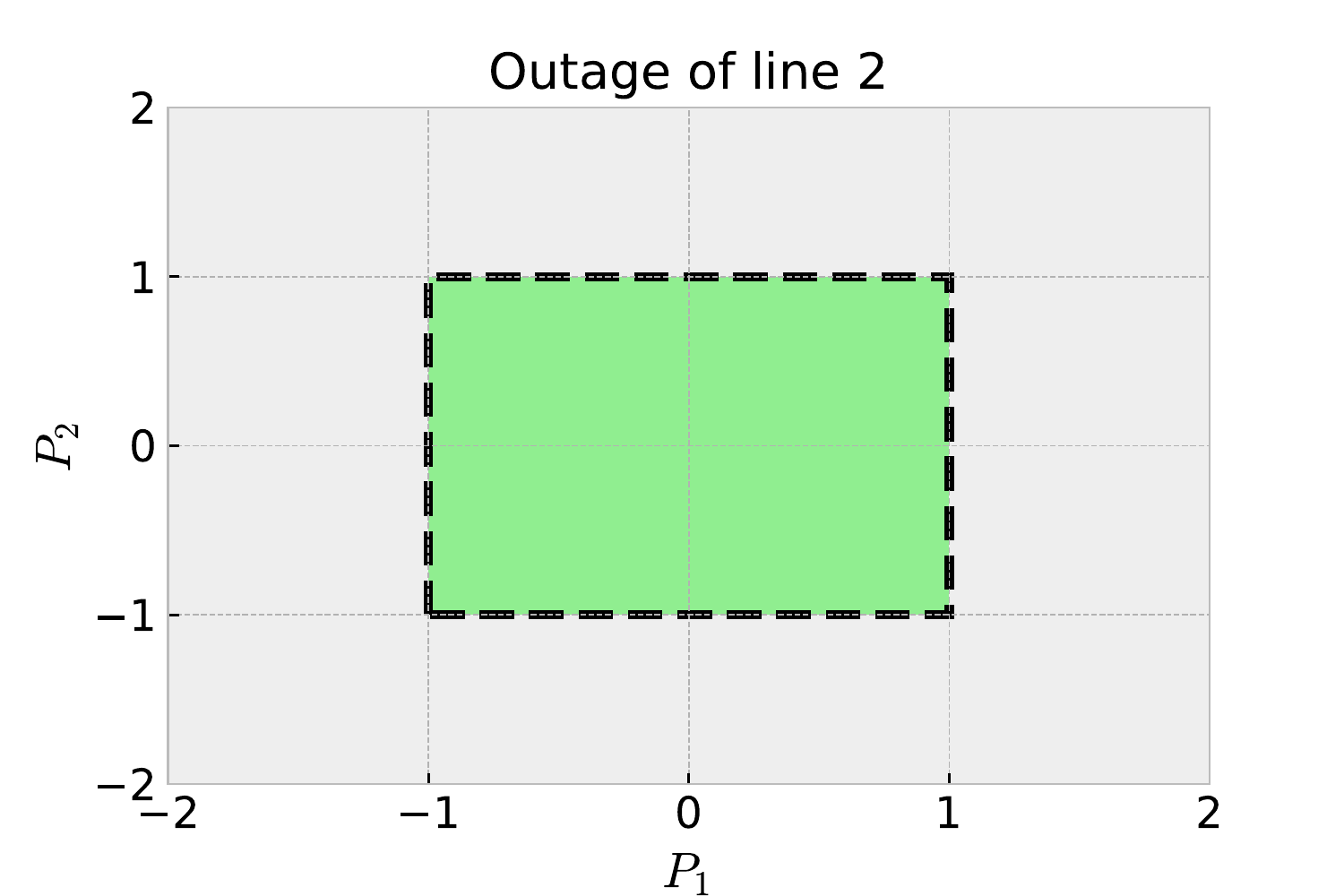}
		\caption{Line 2 failure scenario}
		\label{fig:sub-first}
	\end{subfigure}
	\begin{subfigure}{.4\textwidth}
		\centering
		\includegraphics[width=\linewidth]{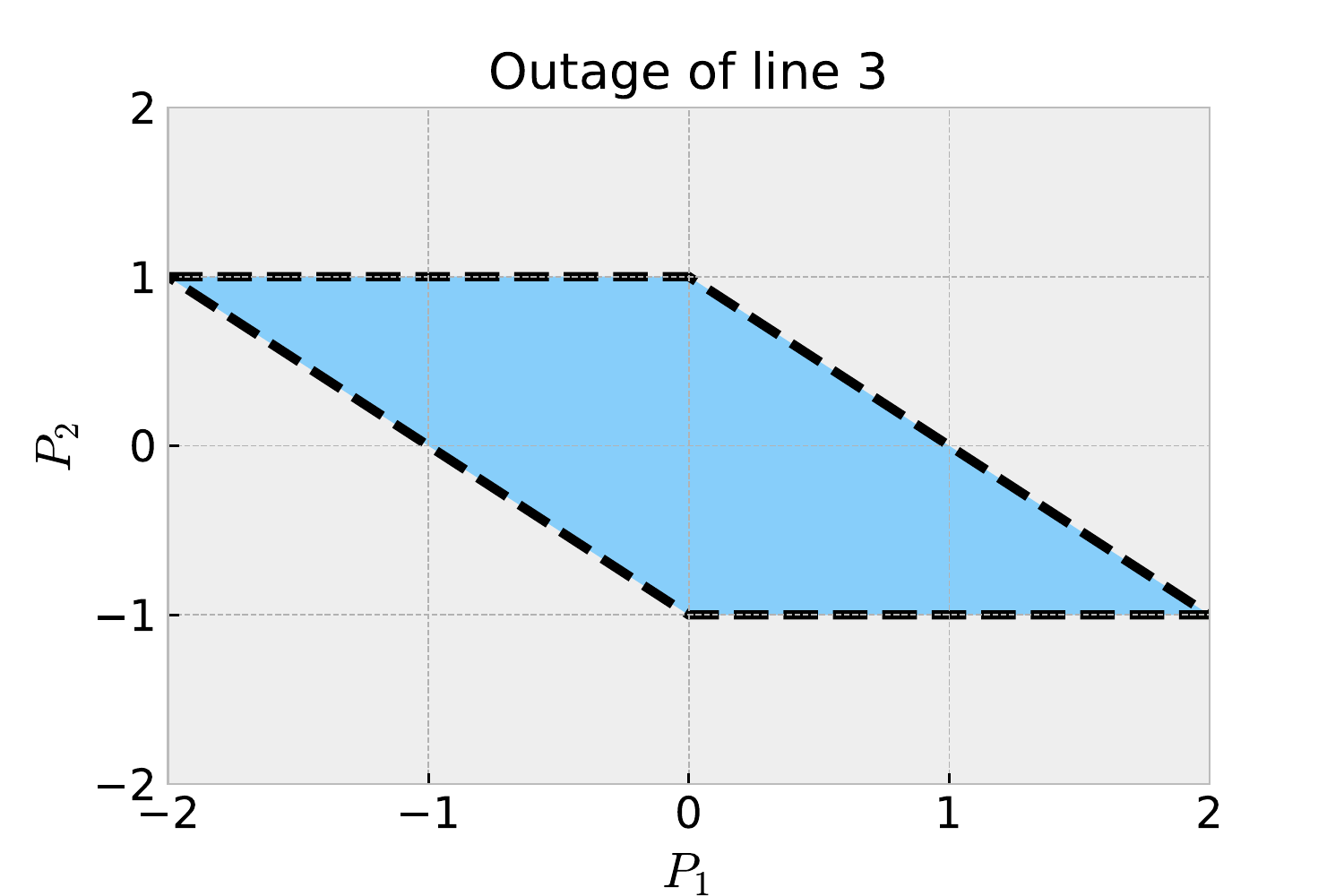}
		\caption{Line 3 failure scenario}
		\label{fig:sub-second}
	\end{subfigure}
	\caption{Polygon of feasible region between $p_{1}$ and $p_{2}$ in a 3-bus
	triangular network.}
	\label{fig:poly_n0_n1}
\end{figure}

\begin{figure}[!t]
	\begin{subfigure}{.4\textwidth}
		\centering
		\includegraphics[width=\linewidth]{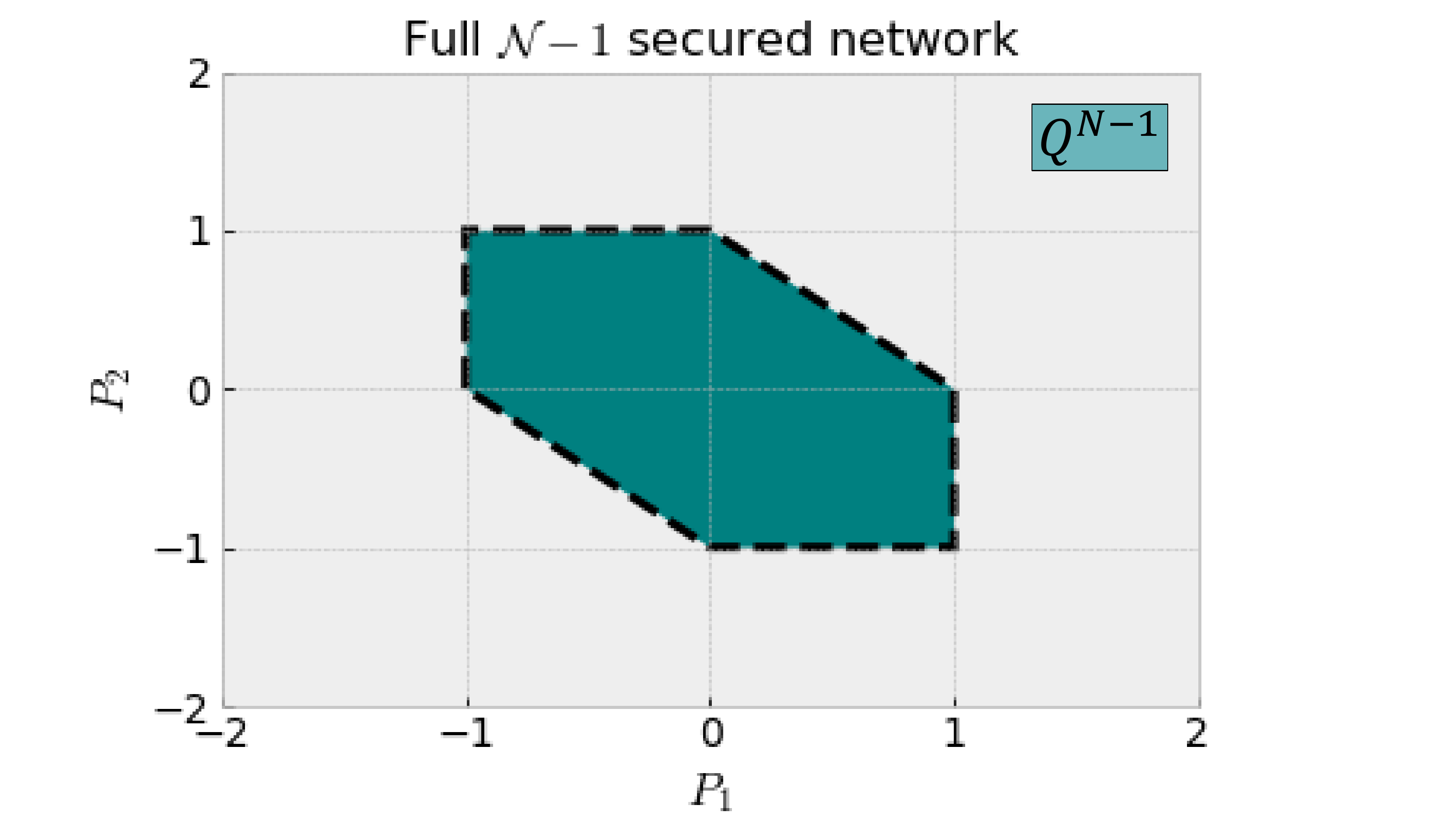}
		\caption{}
		\label{fig:sub-first}
	\end{subfigure}
	\begin{subfigure}{.4\textwidth}
		\centering
		\includegraphics[width=\linewidth]{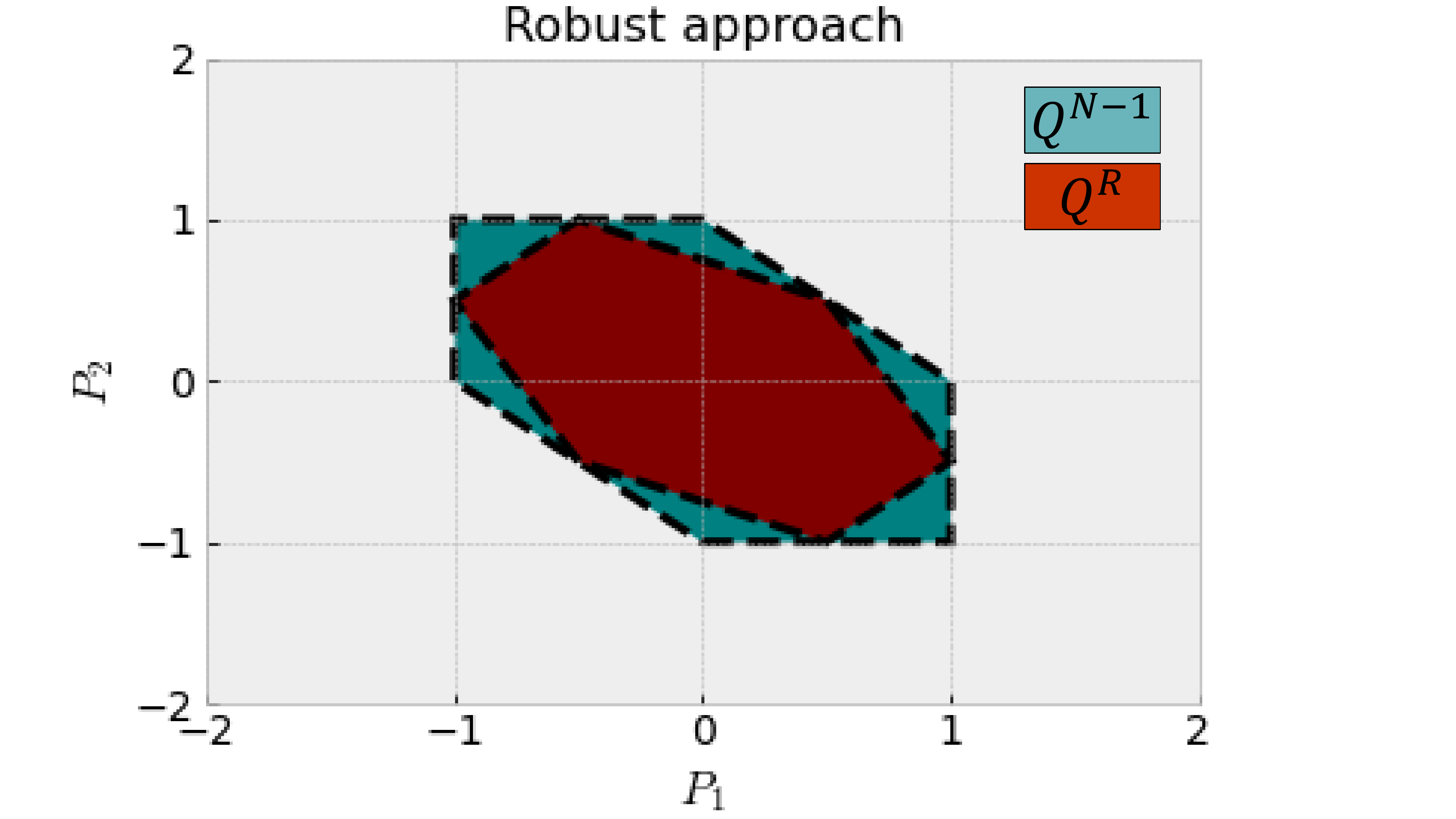}
		\caption{}
		\label{fig:sub-third}
	\end{subfigure}
	\vspace{2mm}
	\begin{subfigure}{.4\textwidth}
		\centering
		\includegraphics[width=\linewidth]{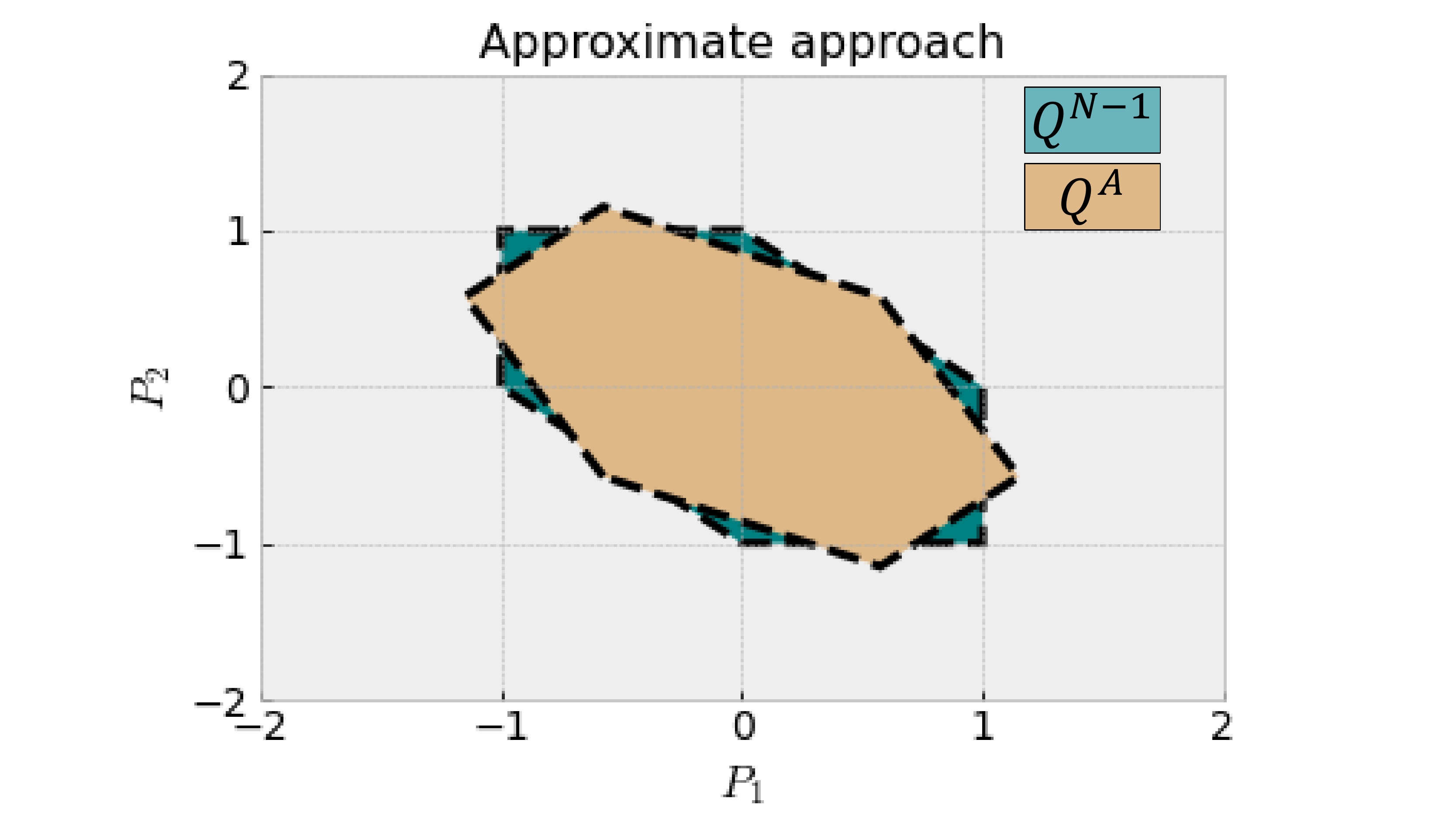}
		\caption{}
		\label{fig:sub-first}
	\end{subfigure}
	\caption{Polygon of feasible region between $p_{1}$ and $p_{2}$: for fully
	secured $\mathcal{N}-1$  network (a), based on the robust approach and fully
	secured $\mathcal{N}-1$ network (b), based on the approximate approach and
	fully secured $\mathcal{N}-1$ network (c) in a 3-bus triangular network.}
	\label{fig:poly_full_n1}
\end{figure}

\section{Simulation results} \label{sec:simulation}

In this section, the performance of our proposed approaches to determine buffer
capacity factors are studied. The LOPF problem is simulated in PyPSA
\citep{PyPSA}. The simulation results of our proposed approaches are evaluated
in different test grids, including a 3-bus triangular network, some small-scale
network topologies and a 50-bus Germany network as a large-scale network derived
from PyPSA-Eur, an open-source model of the European electricity transmission
system \citep{horsch2018pypsa}.

\subsection{3-bus network}

In this section, the performance of our proposed methodologies is studied in a
3-bus triangular network. Fig. \ref{fig:node_n0_n1}a displays the
$\mathcal{N}-0$ scenario when there exist no line outage in the network. The
$\mathcal{N}-1$ scenarios are also illustrated in Fig. \ref{fig:node_n0_n1}b
where only one line outage has occurred in each of the scenarios. Assuming bus
$1$ is a slack bus, there are two nodal net power injections ($p_{1}$ and
$p_{2}$) in the system. Thus, there exists a 2-dimensional polytope (so-called
\textit{polygon}) \citep{toth2017handbook}. Fig. \ref{fig:poly_n0_n1}
demonstrates the feasible region between $p_{1}$ and $p_{2}$ in $\mathcal{N}-0$
and $\mathcal{N}-1$ scenarios, respectively. According to Fig.
\ref{fig:poly_n0_n1}, feasible region between $p_{1}$ and $p_{2}$ for the fully
secured $\mathcal{N}-1$ network is illustrated in Fig. \ref{fig:poly_full_n1}a.
As mentioned in Section \ref{sec:heuap}, we propose three approaches consisting
of robust, approximate and line-specific approaches in this paper. Fig.
\ref{fig:poly_full_n1}c shows the polygons of feasible regions between $p_{1}$
and $p_{2}$ based on the approximate approach and fully secured $\mathcal{N}-1$
network in a 3-bus triangular grid. The total volumes of both polygons, as well
as the volumes of mismatched regions, are equal, as shown in Fig.
\ref{fig:poly_full_n1}c. On the other hand, the polygons of feasible regions
between $p_{1}$ and $p_{2}$ based on the robust approach and fully secured
$\mathcal{N}-1$ network are illustrated in Fig. \ref{fig:poly_full_n1}b. The
polygon that came from the robust approach is completely covered by the polygon
of fully secured $\mathcal{N}-1$ network, and this is a reason why this approach
is called robust. In other words, the feasible region of the robust approach is
a subset of the feasible of the fully secured $\mathcal{N}-1$ network, so
$c^{R}$ (which is not a line-dependant factor) guarantees the reliability of the
system in case of a single-line outage in the transmission network.

\subsection{Small-scale networks}
\label{section:small-eu}

\begin{figure}[!t]
	\centering
	\includegraphics[width=0.8\columnwidth]{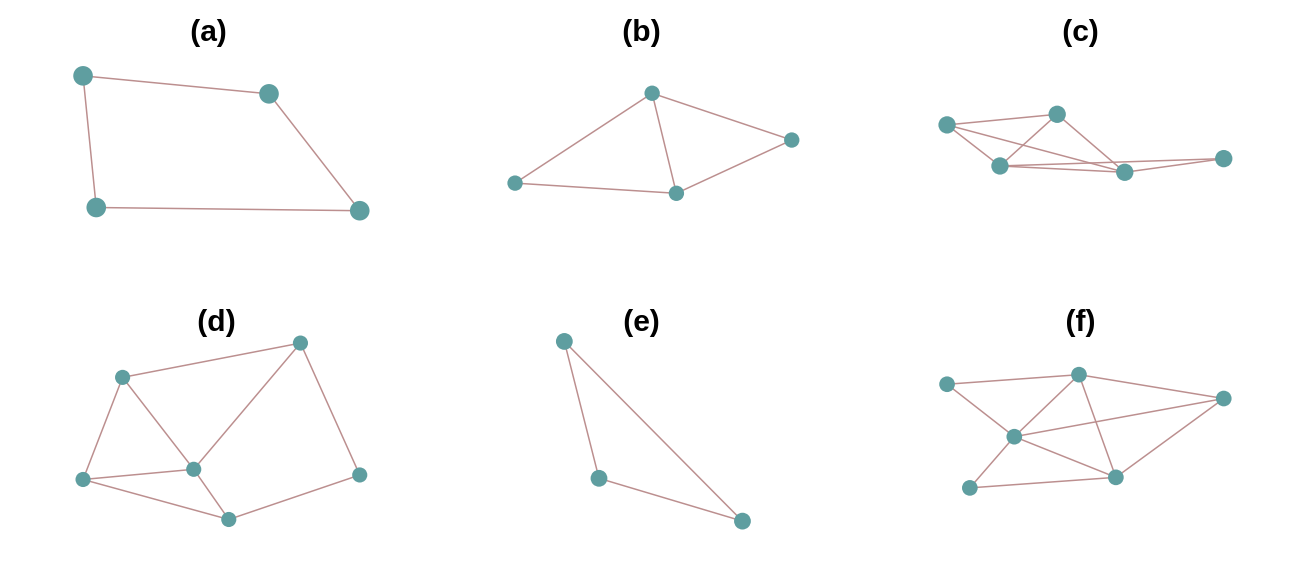}
	\caption{Topology of small-scale networks.}
	\label{fig:eu_small_net}
\end{figure}

\begin{figure}[!t]
	\centering
	\includegraphics[width=0.8\columnwidth]{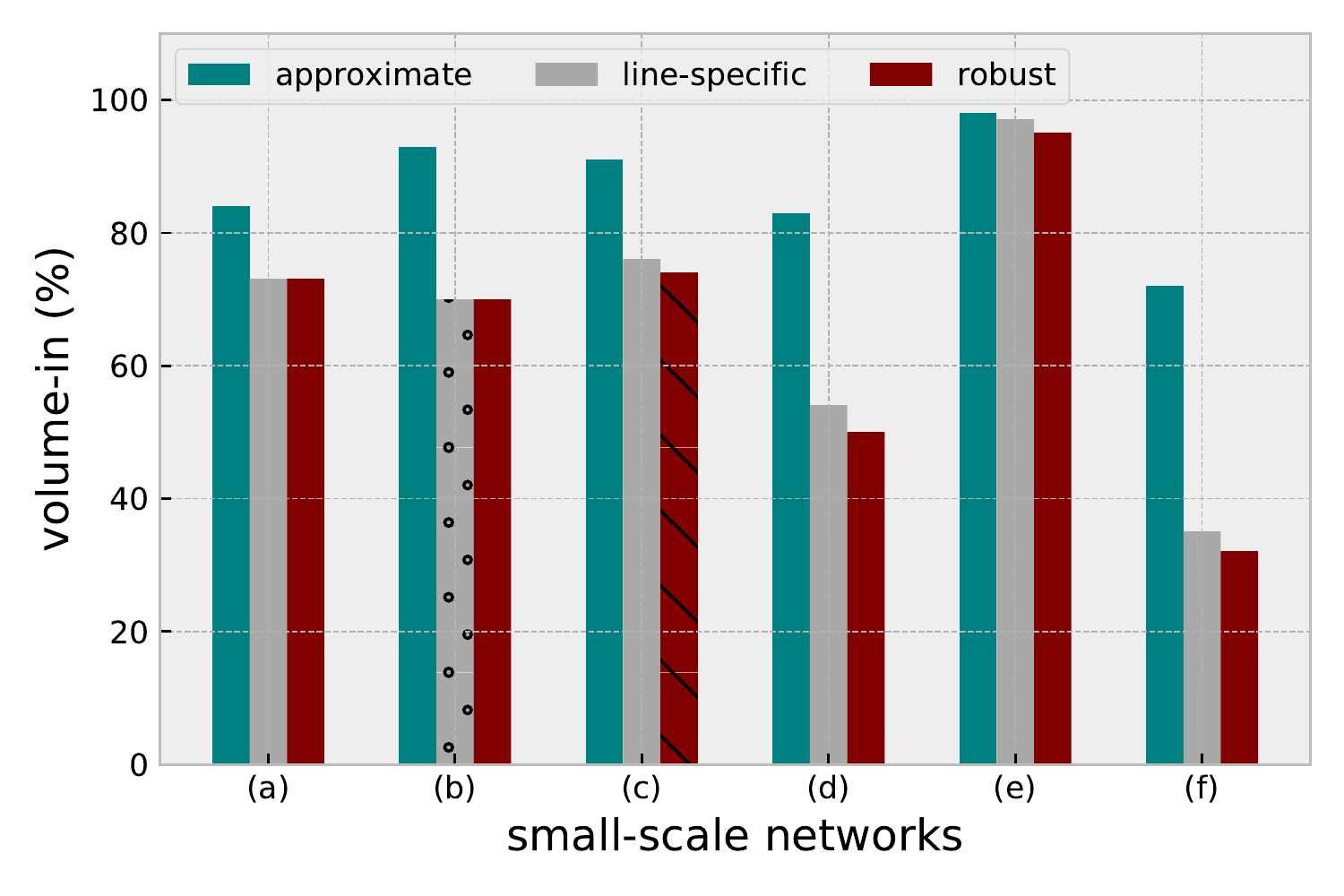}
	\caption{Volume-in ($\Gamma$) of the proposed heuristic approaches in small-scale European countries.}
	\label{fig:eu_small_vol}
\end{figure}

In this section, the performance of the proposed approaches for small-scale
network topologies is evaluated 
European networks, which is  derived from highly aggregated European
transmission network, without cross-border connections with their neighborhoods
as shown in Fig. \ref{fig:eu_small_net}. 
buffer capacity factor of each line obtained by the line-specific for
small-scale European networks.
Table \ref{tab:c_eu_small} presents the values of $c^{A}$ and $c^{R}$ for
small-scale networks. As seen in Table \ref{tab:c_eu_small}, $c^{R}$ is less
than $c^{A}$ in all networks. Moreover, $c^{A}$ is around 0.7 for all
small-scale networks. However, $c^{R}$ is not around 0.7 for all networks which
is because of their specific topology. 
Table \ref{tab:c_eu_small}, $c_{\ell}$ for some of transmission lines are higher
than $c^{R}$. Additionally, the relation between the redundant capacity and
buffer capacity factor is represented in Fig. \ref{fig:cl_tl}. Thus, increment
of $t_{\ell}$ is in line with the increment of $c_{\ell}$. In other words, the
lines with higher redundant capacity could have higher buffer capacity factor as
seen in Fig. \ref{fig:cl_tl}. 
Besides, the volume-in criteria ($\Gamma$), which is defined in Eq.
(\ref{eq:volumin}), obtained by approximate, robust and line-specific
approaches, is displayed in Fig. \ref{fig:eu_small_vol}.

\begin{eqnarray}\label{eq:volumin} \Gamma=\frac{V(Q^{heuristic} \cap
	Q^{\mathcal{N}-1})}{V(Q^{\mathcal{N}-1})}
\end{eqnarray}

According to Fig. \ref{fig:eu_small_vol}, $\Gamma$ is maximum in the approximate
approach. Moreover, $\Gamma$ of the line-specific approach is higher than the
robust one presenting the effectiveness of the line-specific approach. However,
note that even if the volume-in is high for the approximate approach, there will
be parts of $Q^{A}$ outside $Q^{\mathcal{N}-1}$, making some solutions
infeasible for $\mathcal{N}-1$ .

\begin{table}[t]
	\renewcommand{\arraystretch}{2}
	\centering
	\caption{ Robust and approximate buffer capacity factors for small-scale networks.}
	\label{tab:c_eu_small}
	\begin{tabular}{ccc|ccc}
		\hline
		Network & $c^{A}$ & $c^{R}$ & Network & $c^{A}$ & $c^{R}$ \\ \hline
		(a)     & 0.66    & 0.59    & (d)     & 0.71    & 0.62    \\ \hline
		(b)     & 0.84    & 0.74    & (e)     & 0.8     & 0.78    \\ \hline
		(c)     & 0.72    & 0.67    & (f)     & 0.7     & 0.55    \\ \hline
	\end{tabular}
\end{table}

\subsection{Large-scale network: German transmission system}
\label{section:de}

\begin{figure}[!t]
	\begin{subfigure}{.5\textwidth}
		\centering
		\includegraphics[width=0.8\linewidth]{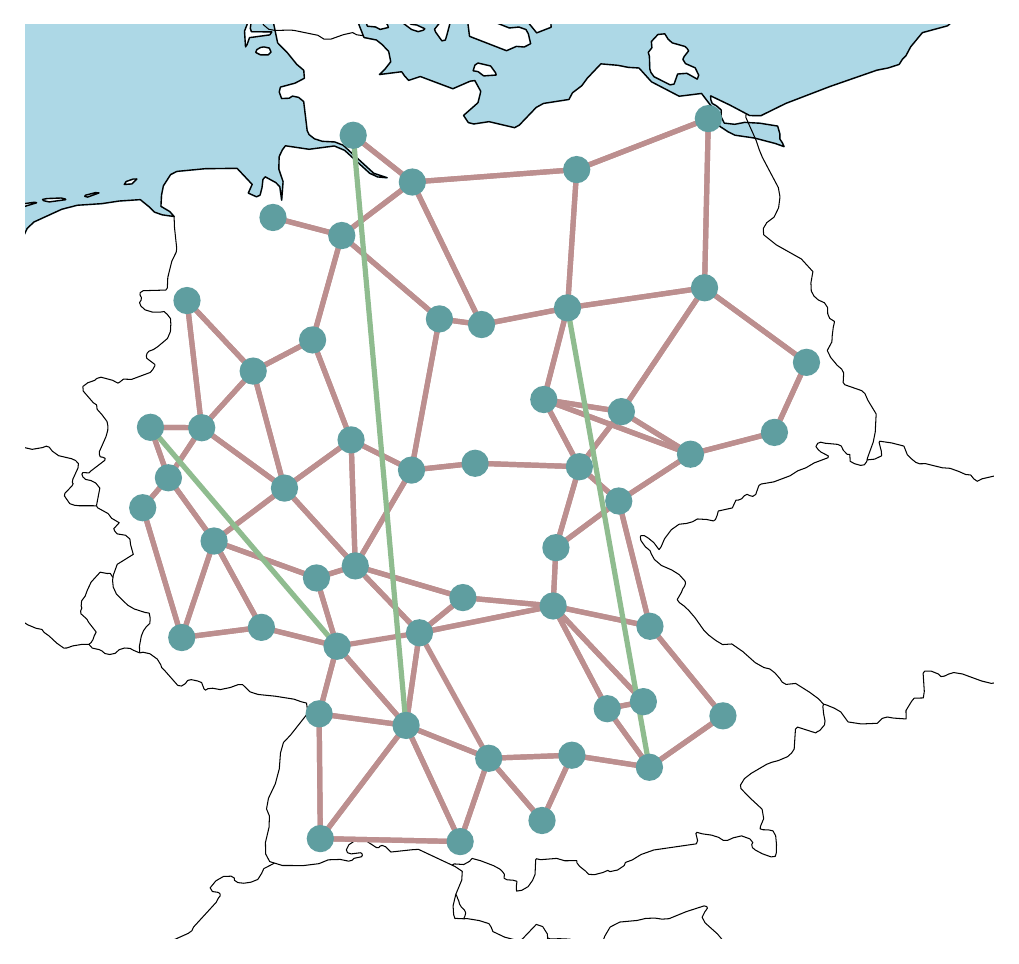}
		\caption{}
		\label{fig:de_net}
	\end{subfigure}
	\begin{subfigure}{.5\textwidth}
		\centering
		\includegraphics[width=\linewidth]{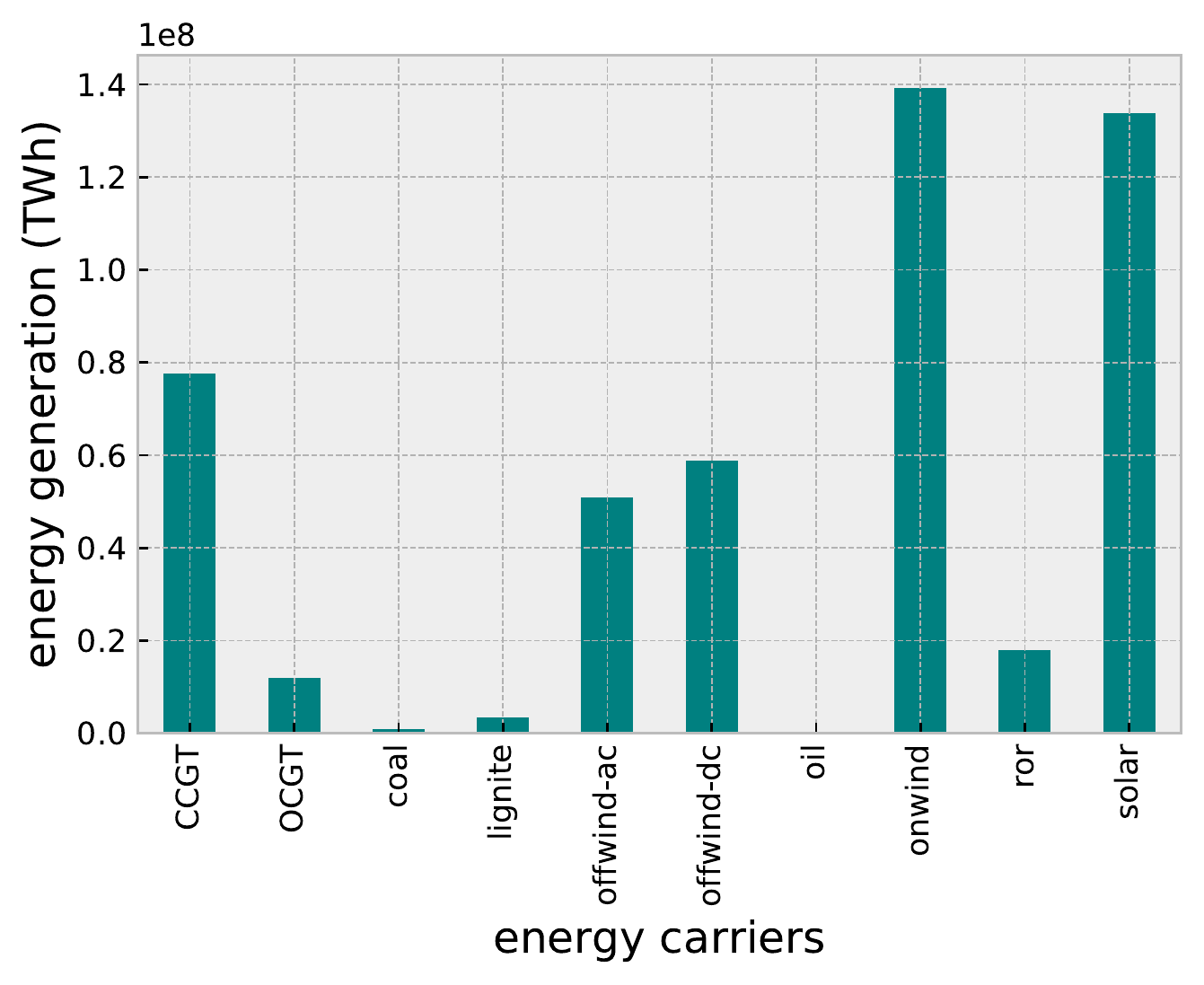}
		\caption{}
		\label{fig:carrier}
	\end{subfigure}
	\caption{German transmission network case study (a) map of German transmission network with 50 buses in which green and red lines show HVDC and HVAC transmission lines, respectively, and (b) yearly energy generation by energy carriers in German transmission network.}
	\label{fig:map_carrier}
\end{figure}

In this section, the buffer capacity factor for the Germany network as a
large-scale case study is found by our proposed methodology for a large number
of buses as presented in Section \ref{sec:large}. Fig. \ref{fig:de_net} displays
50-bus German transmission. Besides, Fig. \ref{fig:carrier} shows the total
annual energy generation of energy carriers in the German transmission network.
Additionally, a $90\%$ carbon emission reduction scenario is considered in this
paper so that  $\mathcal{E}_{CO_2}$ is just 10\% of the electricity sector
emissions in Germany in 1990. Table \ref{tab:c_de} presents buffer capacity
factors for different subsets of the 50-bus Germany network based on the
proposed clustering algorithm. The subsets represented in Table \ref{tab:c_de}
are obtained by the k-means clustering algorithm and based on geographic
proximity. As seen in Table \ref{tab:c_de}, we assume 13 subsets for the 50-bus
Germany network and determine $c^R$ and $c^A$ for each subset.

\begin{figure}[!t]
	\begin{subfigure}{.4\textwidth}
		\centering
		\includegraphics[width=\linewidth]{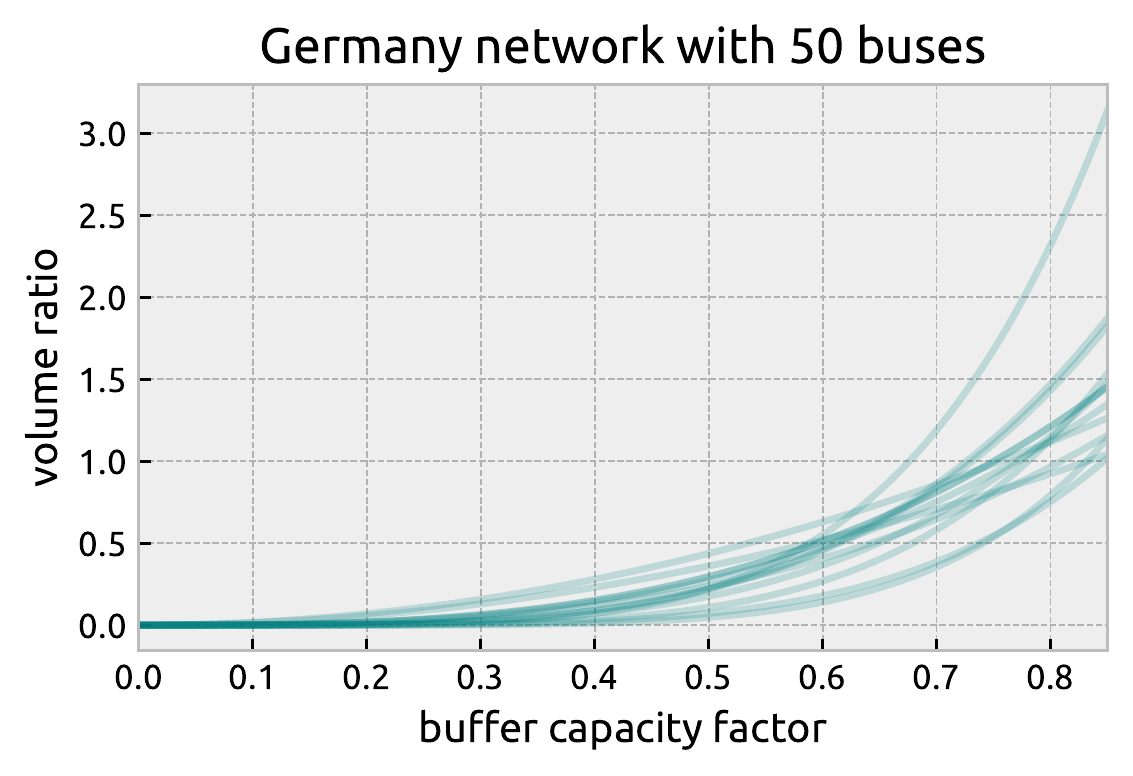}
		\caption{}
		\label{fig:volumin}
	\end{subfigure}
	\begin{subfigure}{.4\textwidth}
		\centering
		\includegraphics[width=\linewidth]{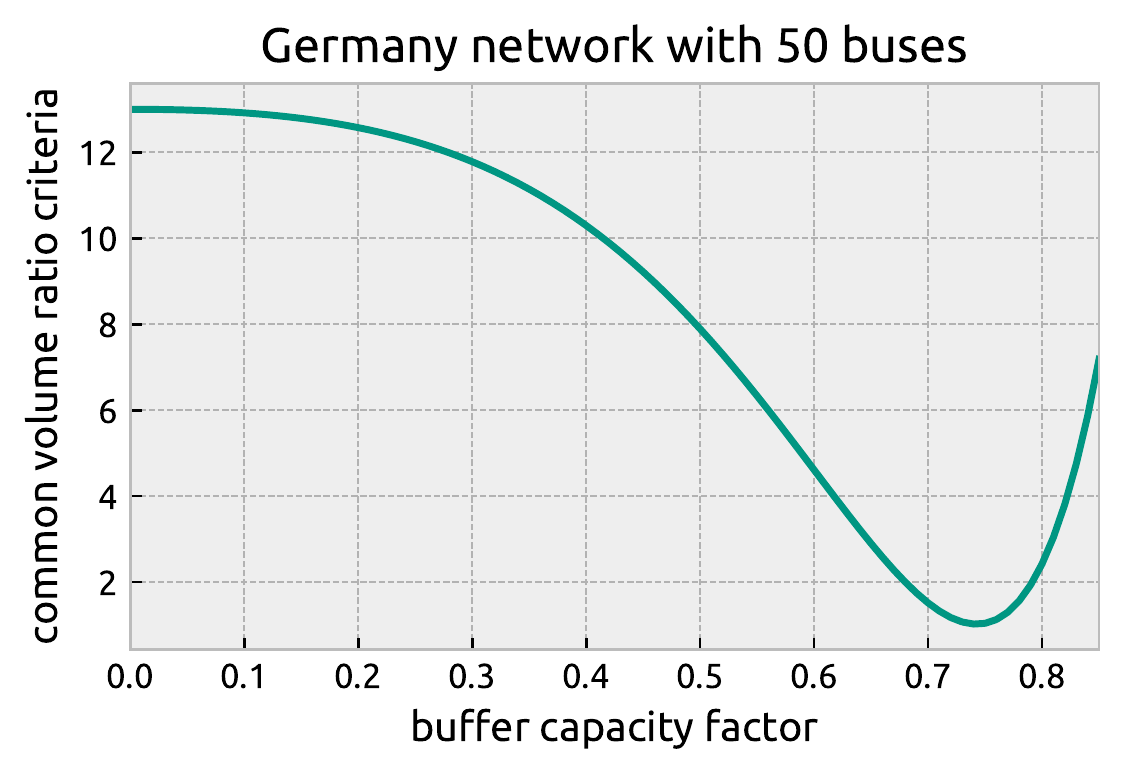}
		\caption{}
		\label{fig:cvi}
	\end{subfigure}
	\caption{Impact of buffer capacity factors based on k-means clustering on (a) volume ratio in Germany network (b) common volume ratio criteria in Germany network.}
	\label{fig:volumin_cvi}
\end{figure}

As seen in Table \ref{tab:c_de}, $c^R$ for the 50-bus Germany network is $0.59$
which is obtained based on Eq. (\ref{eq:crob}) which is the minimum of robust
buffer capacity factor for each subset. On the other hand, Fig.
\ref{fig:volumin} shows the impact of buffer capacity factor on $\Upsilon$ in
all subsets of the Germany network.  As in k-means clustering, each bus belongs
only to one cluster, we miss lines that cross clusters. In other words, the
results came from k-means clustering do not guarantee $\mathcal{N}-1$ security
of our proposed robust approach. Thus, the robust buffer capacity factor is
found by 100 random samples to cover all transmission lines in the transmission
network, and $c^{R}$ is found $0.59$ which confirms outcomes presented by Table
\ref{tab:c_de}.

\begin{table}[!t]
	\renewcommand{\arraystretch}{2}
	\centering
	\caption{ Robust and approximate buffer capacity factors based on k-means clustering for clusters of the 50-bus Germany network.}
	\label{tab:c_de}
	\begin{tabular}{ccc|ccc}
		\hline
		Cluster & $c^{R}$ & $c^{A}$ & Cluster & $c^{R}$ & $c^{A}$ \\
		\hline
		0       & 0.62    & 0.77    & 7       & 0.66    & 0.75    \\
		\hline
		1       & 0.67    & 0.76    & 8       & 0.66    & 0.77    \\
		\hline
		2       & 0.73    & 0.81    & 9       & 0.65    & 0.73    \\
		\hline
		3       & 0.8     & 0.81    & 10      & 0.71    & 0.83    \\
		\hline
		4       & 0.71    & 0.75    & 11      & 0.78    & 0.85    \\
		\hline
		5       & 0.59    & 0.68    & 12      & 0.61    & 0.73    \\
		\hline
		6       & 0.77    & 0.83    & Total   & 0.59    & 0.74    \\
		\hline
	\end{tabular}
\end{table}

\begin{figure}[!t]
	\centering
	\includegraphics[width=0.5\columnwidth]{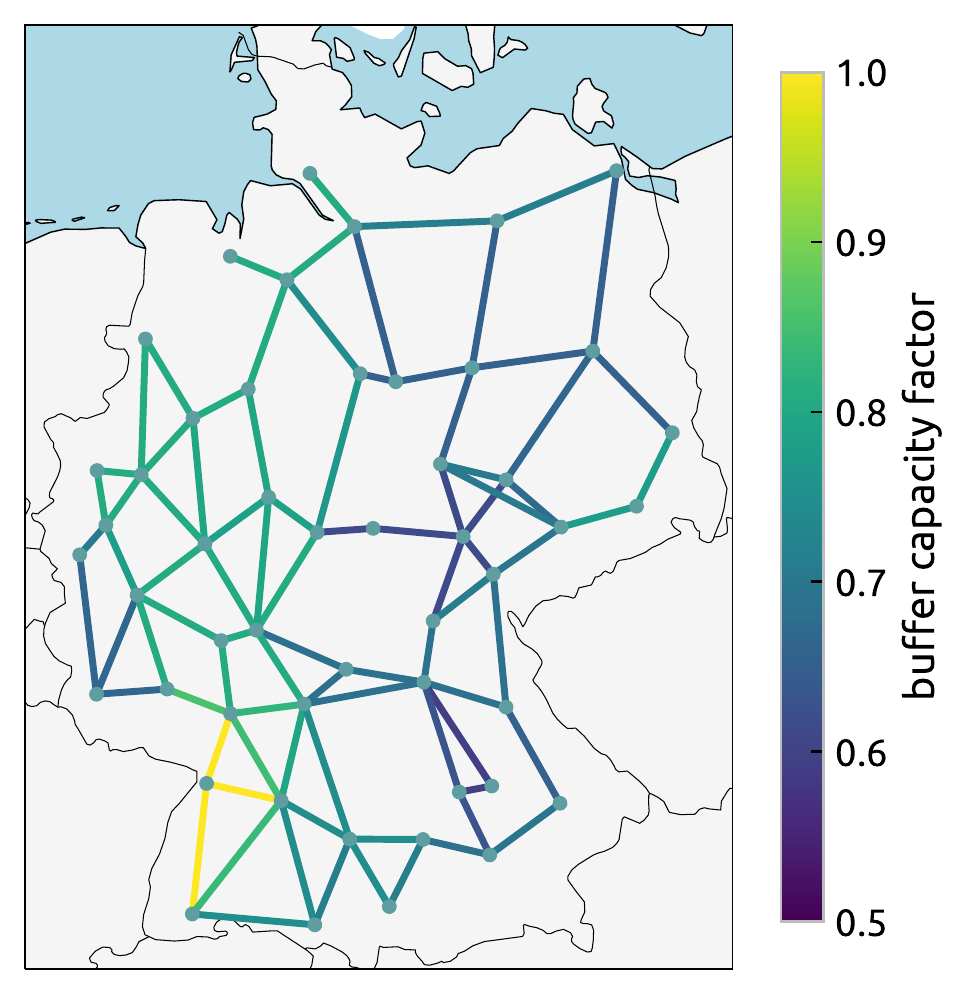}
	\caption{Line-specific buffer capacity factor in Germany network.}
	\label{fig:clinemap}
\end{figure}

\begin{figure}[!t]
	\centering
	\includegraphics[width=.9\columnwidth]{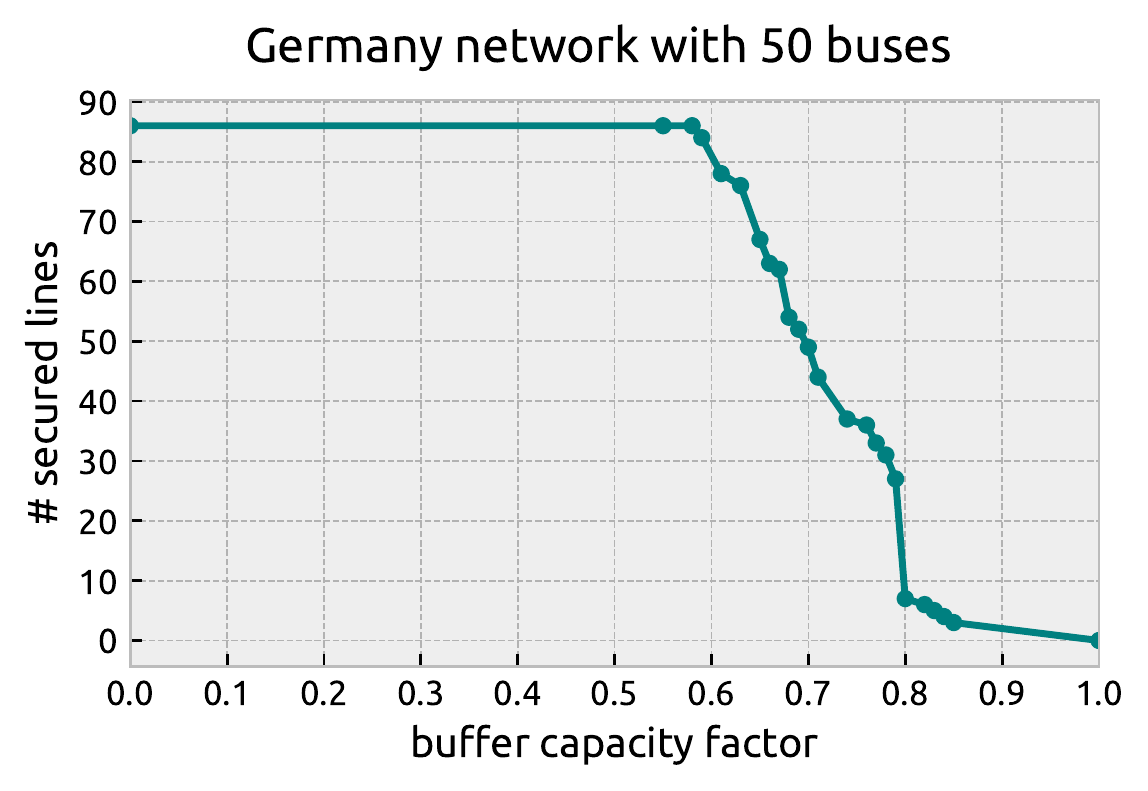}
	\caption{Total number of secure transmission lines with different buffer capacity factors in Germany network.}
	\label{fig:cline}
\end{figure}

As seen in Fig. \ref{fig:volumin}, increment of the buffer capacity factor
increases $\Upsilon$ for all subsets. However, their corresponding increment
patterns are not the same. Moreover, $c^A$ for each subset is different as
presented in Table \ref{tab:c_de}. As shown in Fig. \ref{fig:cvi} and according
to Eq. (\ref{eq:C_A_de}), there exists a minimum in the common volume criteria
curve with buffer capacity equals $0.74$ as presented in Table \ref{tab:c_de}.
Fig. \ref{fig:clinemap} illustrates the map of line-specific buffer capacity
factors in Germany network which is calculated based on Eq. \eqref{eq:C_l_de}.
According to Fig. \ref{fig:clinemap}, $c_\ell$ is in range of $0.59$ and $1$ for
each line of the transmission network where lines with higher $c_\ell$ are more
secure. Additionally, Fig. \ref{fig:cline} shows the impact of buffer capacity
factor on total number of secure transmission lines. As seen in Fig.
\ref{fig:cline}, for $c \leq 0.59$ all transmission lines are secured which
proves $c^R$ equals $0.59$ for the German transmission network.

\begin{figure}[!t]
	\begin{subfigure}{.4\textwidth}
		\centering
		\includegraphics[width=\linewidth]{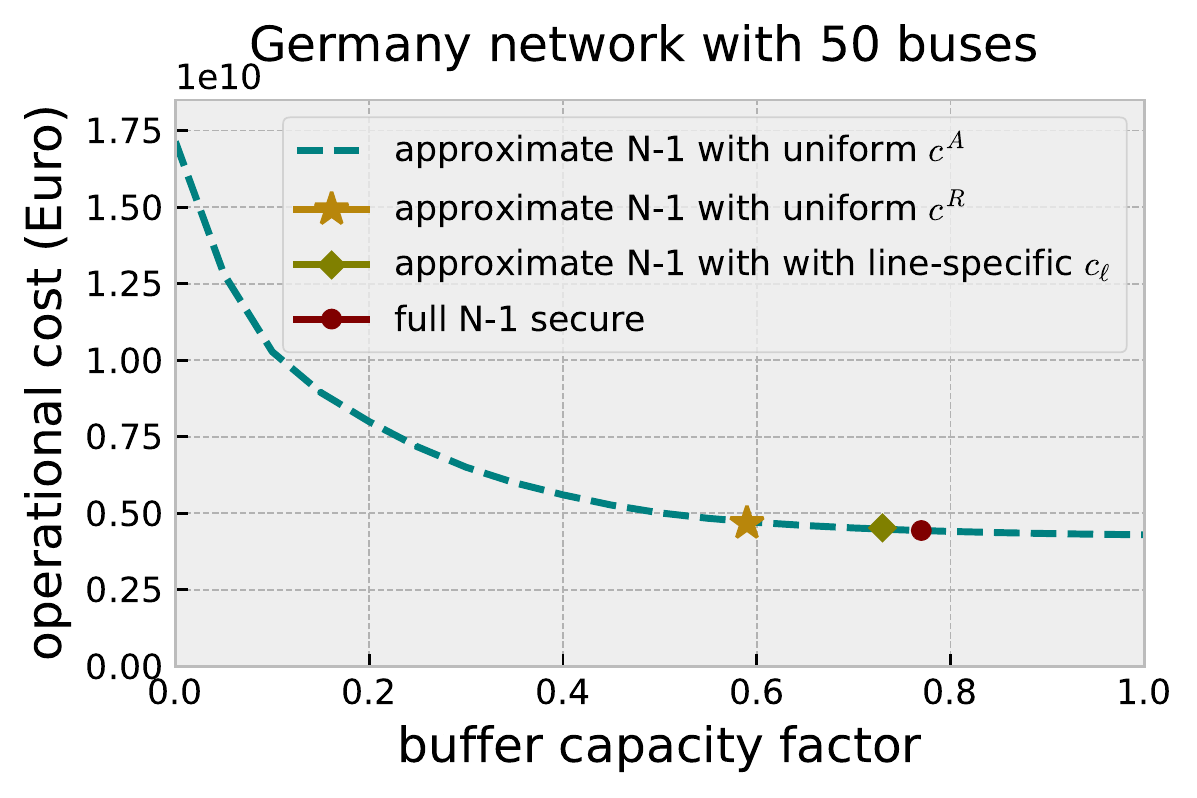}
		\caption{}
		\label{fig:cost}
	\end{subfigure}
	\begin{subfigure}{.4\textwidth}
		\centering
		\includegraphics[width=\linewidth]{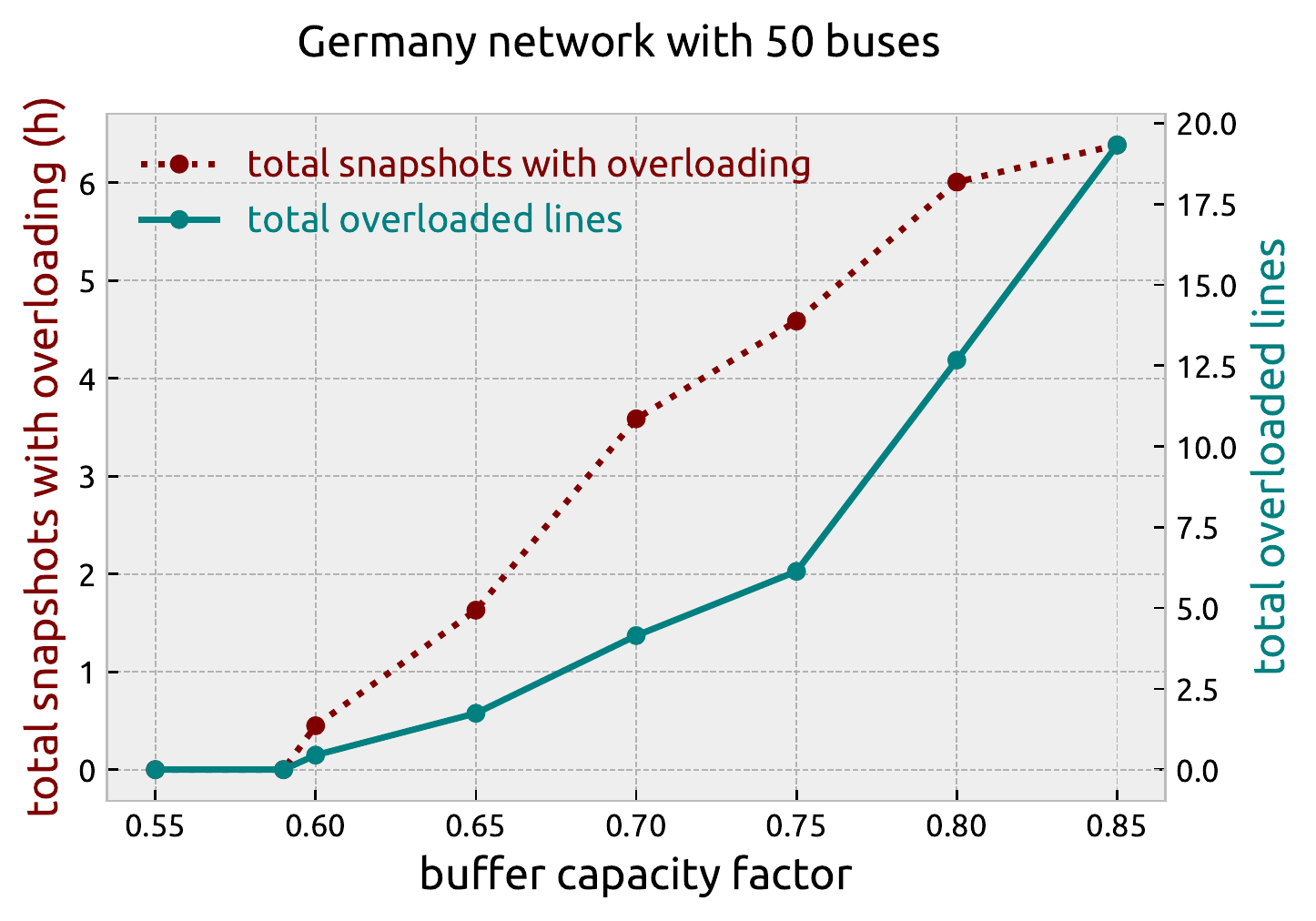}
		\caption{}
		\label{fig:overload}
	\end{subfigure}
	\caption{Impact of buffer capacity factor on (a) system annual operational cost in Germany network (b) total snapshots with overloading and overloaded lines  in Germany network.}
	\label{fig:cost_overload}
\end{figure}

Additionally, the economic impact of the different buffer capacity factor
approaches is studied in the German transmission network. In this analysis,
operations are optimized for a full year of weather and load data for the 50
buses of the Germany network, assuming a generation fleet that yields a 90\%
CO$_2$ reduction compared to 1990. This leads to a $78\%$ penetration of wind
and solar power generation based on Fig. \ref{fig:carrier}. Fig. \ref{fig:cost}
illustrates the impact of the buffer capacity factor on the yearly operational
cost in Eq. \eqref{eq:lopf_obj}, in the 50-bus Germany network. According to
Fig. \ref{fig:cost}, increasing the amount of the buffer capacity factor
decreases the operational cost of the system. Moreover, the operational cost of
the system in the fully secured $\mathcal{N}-1$ network is equal to the
operational cost of the system with buffer capacity in the range of $[0.75,
0.8]$ as seen in Fig. \ref{fig:cost}. As the approximate buffer capacity factor,
$c^A$, is found to be $0.74$ in Section \ref{section:de}, it can be concluded
that the operational costs of the system in the fully secured $\mathcal{N}-1$
network is close to the operational cost of the system with $c^A$ which confirms
the intuitions from the industry that a buffer capacity factor of $0.7$ is
suitable \citep{siemens}. It is also noticeable that the close operational
system costs do not necessarily imply that the operational results are the same.
Moreover, Fig. \ref{fig:cost} illustrates that the operational cost based on our
proposed heuristic approach with the line-specific approach is higher than the
operational cost in the fully secured $\mathcal{N}-1$ case and less than the
proposed robust approach. Furthermore, it has been found that operational costs
of the system for a fully secured $\mathcal{N}-1$, approximate $\mathcal{N}-1$
with line-specific buffer capacity factors, and approximate $\mathcal{N}-1$ with
approximate buffer capacity factor are very close. On the other hand, it is
noteworthy that the approximate buffer capacity does not guarantee security of
all transmission lines, whereas all lines are secured based on line-specific
buffer capacity factor. Hence, the line-specific approach is the best approach
among our proposed approaches to approximate the fully secured $\mathcal{N}-1$
conditions in the network, while using uniform approximate buffer capacity
factor can only provide approximation of operational costs of the fully secured
$\mathcal{N}-1$.

\begin{table}[!t]
	\renewcommand{\arraystretch}{1}
	\centering
	\caption{ Computational time and memory usage of $\mathcal{N}-0$ problem
	with uniform/line-specific buffer capacity factors and $\mathcal{N}-1$ fully
	secured problem.}
	\label{tab:time_mem}
	\begin{tabular}{ccc}
		\hline
		                                   & Computational & Memory        \\
		                                   & time [Sec]    & usage [MB]    \\
		\hline
		$\mathcal{N}-0$ with uniform       & $\approx$ 270 & $\approx$ 860 \\
		buffer capacity factor             &               &               \\
		\hline
		$\mathcal{N}-0$ with line-specific & 326           & 860           \\
		buffer capacity factor             &               &               \\
		\hline
		$\mathcal{N}-1$ fully secured      & 3786          & 5582          \\
		\hline
	\end{tabular}
\end{table}

According to Fig. \ref{fig:cost}, the robust case that guarantees no overloading
is $4\%$ more expensive than the fully secured $\mathcal{N}-1$ case since its
feasible space is smaller. Moreover, the operational cost of the line-specific
approach is $2\%$ lower than the robust one. Thus, while line-specific and
robust approaches guarantee no overloading caused by a single-line outage in the
network, the operational cost of the line-specific case is less than the robust
one showing the benefit of the line-specific approach. Table \ref{tab:time_mem}
presents the computational time and memory usage of our proposed heuristic
approaches with the uniform (approximate and robust) and line-specific buffer
capacity factors in comparison with the fully secured $\mathcal{N}-1$ problem.
As seen in Table \ref{tab:time_mem}, the computational time and memory usage in
our proposed method is significantly less than the fully secured case, which is
because of fewer constraints in our proposed approximate $\mathcal{N}-1$ case.
Moreover, Fig. \ref{fig:overload} shows total cases and snapshots where the
transmission network is overloaded in at least one line for different buffer
capacity factors. As shown in Fig. \ref{fig:overload}, there exists no
overloading in the system for the buffer capacity factor less than $0.59$ (as
$c^R$ has been obtained $0.59$ according to Fig. \ref{fig:cline} and Table
\ref{tab:c_de}). In other words, Fig. \ref{fig:overload} proves that the
transmission network with a robust buffer capacity factor is guaranteed to be
fully secured, i.e. no transmission line is ever overloaded.

\section{Conclusion}
\label{sec:conclusion}

In this paper, we have introduced a novel heuristic method for modeling the
fully secured network considering $\mathcal{N}-1$ contingencies based on the
topology of the transmission network in the security-constrained linearized AC
optimal power flow problem. To this end, the network's secure feasible space has
been determined by comparing the polytopes of the feasible region among nodal
net power. Additionally, secure feasible regions have then been approximated
using far fewer constraints by using the $\mathcal{N}-0$ constraints with buffer
capacity factors that have been obtained by line-specific, robust and
approximate approaches to constrain the maximum loading of lines. The algorithms
have been applied on smaller subsets and clusters of the buses to find
line-specific, robust and approximate buffer capacity factors for networks with
many buses. According to the simulations results of the test networks, it has
been found that:

\begin{itemize}
	\item The industry heuristic of $0.7$ buffer capacity is roughly suitable
	for the Germany network to provide fast results (since there are fewer
	constraints) while not producing significant overloading, and giving a
	system cost similar to the actual $\mathcal{N}-1$ case.
	\item For absolute robustness while having fast results, you need to shrink
	the feasible space smaller than the true $\mathcal{N}-1$ polytope, which
	leads to overestimation of the system cost. This can be mitigated by a novel
	line-specific buffer capacity, which better estimates the shape of the
	$\mathcal{N}-1$ polytope while remaining robust.
	\item These speed-ups are particularly important when we include investment
	optimisation, which tends to increase the computational burden. Including
	fully secured $\mathcal{N}-1$ constraints while optimising over a full year
	to dimension generation, storage, and transmission assets would be
	impossible, so these heuristics provide a good alternative that is still
	accurate.
\end{itemize}

In the future, maintaining $\mathcal{N}-1$ security preventatively can be
superseded with methods that use distributed flexibility to deal with
contingencies like line failures reactively as they arise, thus allowing network
assets to be fully used in regular operation.

\section*{Acknowledgement}

This paper was conducted as part of the CoNDyNet2 project, which was supported
by the German Federal Ministry of Education and Research under grant number
03EK3055E.

\bibliographystyle{elsarticle-num.bst}
\bibliography{bib}

\end{document}